\documentclass{emulateapj}

\shorttitle{The sensitivity of core-collapse supernovae to nuclear electron capture}
\shortauthors{Sullivan et al.}
\def\nuc#1#2{{}^{#1}\mathrm{#2}}

\usepackage{ulem}

\usepackage{amsmath}

\usepackage{ifpdf}
\ifpdf
  \usepackage{epstopdf}
  \usepackage{microtype}
  \microtypesetup{%
	protrusion=true,
	expansion=true,
	tracking=smallcaps,
	letterspace=0,
	tracking=true,
	kerning=true,
	spacing=true,
	babel=true,
	final=true,
  }
  \microtypecontext{spacing=nonfrench}
  \UseMicrotypeSet[protrusion]{alltext}
\else
  \usepackage[dvips]{graphicx}
\fi

\usepackage{hyperref}
\hypersetup{%
	plainpages=false,%
	hypertexnames=true,%
	colorlinks,%
	breaklinks,%
	citecolor=linkblue,%
	linkcolor=linkblue,%
	filecolor=linkbrown,%
	urlcolor=linkblue,%
	pdfcreator={pdfTeX 3.1415926-2.5-1.40.14 (TeX Live 2013)},%
	pdffitwindow=true,
	pdfauthor={Chris Sullivan},%
	pdfkeywords={},%
	pdfpagelayout=TwoPageLeft,
	pdfdisplaydoctitle=true,
	pdfcenterwindow=true,
	pdflang={English},
	bookmarksnumbered=true,%
	bookmarksopen=true,%
	verbose=true,
	breaklinks=true,
        linktocpage=true,
	unicode
}
\usepackage{color}
\definecolor{linkblue}{rgb}{0,0.2,0.6}
\definecolor{linkgreen}{rgb}{0,.5,0}
\definecolor{linkbrown}{rgb}{.6,0,0}
\definecolor{linkred}{rgb}{.7,0,0}

\usepackage{amsmath}
\usepackage{bookmark}
\usepackage{booktabs}
\usepackage[flushleft]{threeparttable}
\usepackage{here}
\usepackage[english]{babel}
\usepackage[english]{varioref}

\usepackage{fancyhdr}

\usepackage{ amssymb }

\long\def\/*#1*/{}

\def\nuc#1#2{{}^{#1}\mathrm{#2}}

\begin{document}
\pagenumbering{arabic}
\setcounter{page}{1}

\title{The sensitivity of core-collapse supernovae to nuclear electron capture}
\author{Chris Sullivan\altaffilmark{1,2,3}$^\dagger$, Evan O'Connor\altaffilmark{4,5}\footnotemark[$\ddagger$], Remco G. T. Zegers\altaffilmark{1,2,3}, Thomas Grubb\altaffilmark{1,2,3}, Sam M. Austin\altaffilmark{1,2,3}}
\affil{$^1$National Superconducting Cyclotron Laboratory, 
  Michigan State University, East Lansing, MI 48824, USA}
\affil{$^2$Department of Physics and Astronomy, 
  Michigan State University, East Lansing, MI 48824, USA}
\affil{$^3$Joint Institute for Nuclear Astrophysics: CEE, 
  Michigan State University, East Lansing, MI 48824, USA}
\affil{$^4$North Carolina State University, Department of Physics, Campus Code 8202, Raleigh, NC 27695, USA}
\affil{$^5$CITA, Canadian Institute for Theoretical Astrophysics, Toronto, Canada, M5S 3H8}

\date{\today}
\email{$^\dagger$sullivan@nscl.msu.edu}
\footnotetext[$\ddagger$]{Hubble Fellow}
\begin{abstract}
A weak-rate library aimed at 
investigating the sensitivity of astrophysical environments to variations of 
electron-capture rates on medium-heavy nuclei has been developed. 
With this library, the sensitivity of the core-collapse and early post-bounce phases of 
core-collapse supernovae to nuclear electron-capture is examined by systematically and 
statistically varying electron-capture rates of individual nuclei. The rates are 
adjusted by factors consistent with uncertainties indicated by comparing theoretical 
rates to those deduced from charge-exchange and $\beta$-decay measurements. To ensure a model 
independent assessment, sensitivity studies across a comprehensive set of progenitors and 
equations of state are performed. In our systematic study, we find a +16/-4\,\% 
range in the mass of the inner-core at the time of shock formation and a $\pm 20\%$ range of peak 
$\nu_e$-luminosity during the deleptonization burst. These ranges are each five times as large as 
those seen from a separate progenitor study in which we evaluated the sensitivity of these parameters 
to 32 presupernova stellar models. It is also found that the simulations are more 
sensitive to a reduction in the electron-capture rates than an enhancement, and in particular 
to the reduction in the rates for neutron-rich nuclei near the $N=50$ closed neutron-shell. 
As measurements for medium-heavy ($A>65$) and neutron-rich nuclei are sparse, 
and because accurate theoretical models which account for nuclear structure considerations 
on the individual nucleus level are not readily available, rates for these nuclei may be 
overestimated. If more accurate estimates confirm this, results from this study 
indicate that significant changes to the core-collapse trajectory can be expected. 
For this reason, experimental and theoretical efforts should focus in this region of 
the nuclear chart.
\end{abstract}

\bookmarksetup{
  italic=false,
  bold=true,
  color=[rgb]{0,.2,.6},
}
\maketitle

\enlargethispage{1cm} {

\section{Introduction}

The study of interactions mediated via the weak nuclear force is of importance to a 
large number of fields in physics. However, it is of particular importance to 
the field of astrophysics because of the longer timescale on which weak 
interactions operate as compared to the strong and electromagnetic interactions.
This is evidenced by the impact that new 
insights into weak reaction physics have on astrophysical models
~\citep{Langanke2003Rev}. Specifically, electron-capture 
reactions play a prominent role in high-density environments such as 
those found in the late stages of massive star evolution~\citep{Heger2001,Martinez-Pinedo2014}, 
thermonuclear~\citep{Iwamoto1999,Brachwitz2000} 
and core-collapse supernovae (CCSNe)~\citep{Hix2003,Janka200738}, neutron 
stars~\citep{Gupta2007,Schatz2014}, and compact object merger events
~\citep{Goriely2015}. Realistic simulations of these environments rely on 
accurate nuclear physics inputs including electron-capture rates.

Electron-capture rates depend sensitively on Gamow-Teller (GT) 
transition-strength distributions in the $\beta^{+}$ direction. These 
transition strengths characterize nuclear excitations in which 
a single unit of spin and isospin are transferred 
($\Delta S, \Delta T=1$), with no transfer 
of orbital angular momentum ($\Delta L=0$). 
While the main component of electron capture occurs on the 
ground state configuration of a nucleus, in high temperature 
stellar environments electron captures on 
thermally-populated excited states of the parent nucleus 
can also contribute significantly to the overall rate 
\citep{Langanke2000}. Unfortunately, it is difficult to obtain information 
about transitions from excited states in the laboratory. 
Compounding the problem is the fact that in order to accurately include 
electron capture in simulations one must include electron captures on a 
wide range of nuclei.
Hence, in general one must rely on theoretical models for a complete description of 
stellar electron-capture rates. 
On the other hand, measurements of Gamow-Teller strength distributions in a 
representative set of nuclei are important for the development and benchmarking 
of robust theories. At the same time, it is critical that 
theoretical and computational efforts provide guidance to experimenters 
on which measurements to perform.

Presently, configuration-interaction (shell-model) calculations are the primary method for 
producing reliable GT strength distributions near stability in the 
$sd$- and $pf$- shells ($8<[N,Z]<20$ and $20<[N,Z]<40$, respectively) 
for electron capture on both ground and excited 
states~\citep{Oda1994,Langanke2003}. Quasi-particle random-phase 
approximation (QRPA) calculations have also been utilized to 
estimate GT strengths for large sets of nuclei, but only 
where transitions from the ground state are considered
~\citep{NPaar2009,Moller19901,Dzhioev2010,YNiu2011,Nabi2004237}. 
Furthermore, comprehensive sets of electron-capture rates (as a function of density 
and temperature) for a large number of nuclei based on QRPA calculations have not 
been published.

Direct and indirect experiments, such as $\beta$-decay and charge-exchange 
(CE) measurements respectively, provide robust benchmarks for theoretical GT 
strengths and therefore are crucial for our understanding of astrophysical 
electron-capture rates. Unfortunately, electron-capture and $\beta$-decay experiments can 
only access states in a limited Q-value window. Furthermore, for neutron-rich nuclei 
$\beta$-decay only provides information in the $\beta^-$ 
direction, which is of limited use for electron-capture studies. 
Intermediate energy ($\gtrsim$ 100 MeV/u) CE reactions in the 
$\beta^+$ direction, however, connect the same initial and final states as electron 
capture, providing information about transitions up to high excitation energies, 
and are thus well suited to study the full Gamow-Teller strength distribution 
of interest. At these energies, CE measurements are accurate at the $\sim$10\% 
level and are therefore able to provide rigorous tests 
of theoretical Gamow-Teller strengths and derived electron-capture rates.

Recently, the results from (n,p), ($d$,$\nuc{2}{He}$), and ($t$,$\nuc{3}{He}$) 
CE reactions on nuclei in the $pf$-shell  
were systematically compared \citep{Cole2012,Scott2014,Noji2014} 
with shell-model calculations using the KB3G \citep{KB3G} and GXPF1a 
\citep{GXPF1A} effective interactions in the $pf$-model space, and with 
calculations based on the QRPA formalism of \citet{Moller19901}. 
The authors compared shell-model and QRPA derived electron-capture 
rates against those derived from CE measurements. 
It was found that the QRPA calculations 
systematically overestimate the electron-capture rates ($\sim$100-3000\%, 
depending on density and temperature), whereas the shell-model estimates produce rates 
similar to those measured experimentally ($\sim$1-50\%)~\citep{Cole2012}. 
Unfortunately, shell-model calculations are 
computationally challenging for nuclei beyond the 
$pf$-shell, and therefore weak rates used in high-density astrophysical 
calculations most commonly rely on less accurate methods. 
In each of these cases, systematic 
and random error exist, and it is therefore 
important to understand the sensitivity of astrophysical simulations 
to uncertainties in these rates. 

Sensitivity studies are useful tools for guiding theoretical and 
experimental efforts because they highlight nuclei 
that should be given particular focus, and they indicate the accuracy 
with which the parameters of interest need to be known. 
They also illustrate how strongly the current parameter 
uncertainties affect the outcome of 
the astrophysical simulations. In this work, we perform $\sim150$ 
collapse simulations examining how systematic and statistical variations of the electron-capture 
rates impact the collapse, bounce and pre-explosion phases of core-collapse 
supernovae simulations over a range of presupernova progenitors and equations 
of state (EOS). We describe the development of a 
modular and open-source weak reaction rate library for use in 
astrophysical simulations and its first implementation using the stellar 
core-collapse code \texttt{GR1D}~\citep{OConnor2010}.
In the following sections, we show that the inner core of the protoneutron star (PNS)
and the observable peak neutrino-luminosity from core bounce and shock formation 
depend sensitively (+16/-4\,\% and $\pm 20$\%, respectively) on the electron-capture 
rates of neutron-rich nuclei. 
As variations on this level are not easily reproduced from uncertainties in other inputs 
to the simulations, they motivate the development of new theoretical models for 
electron-capture rates as well as relevant measurements, which together will constrain 
these and other key parameters discussed in this work.

In Section~\ref{weakrates} we discuss previous development of 
weak rates for astrophysics, as well as the implementations made in 
this work. In Section~\ref{core-collapse} we motivate the importance of 
electron-capture rates during the core-collapse phase of CCSNe, and in 
Section~\ref{codes} we describe the codes we have utilized and developed 
for this work. We detail the sensitivity studies performed in 
Section~\ref{sensitivity} and conclude in Section~\ref{conclusion}.

\section{Astrophysical weak interaction rates}
\label{weakrates}
From the birth to the death of massive stars, the weak nuclear force 
is a primary actor in the story of stellar evolution~\citep{Langanke2003Rev}. 
Weak interactions are important ingredients for nucleosynthesis and also 
for the internal structure of evolving stars, as they sensitively determine the 
electron-to-baryon ratio $Y_e$ and the iron-core mass 
just prior to core-collapse~\citep{Heger2001}. 
Unlike the conditions present during quasistatic stellar evolution, however, in 
the core of a collapsing star the density and temperature are high 
enough that nuclear and electromagnetic reactions equilibrate~\citep{Iliadis2007}. 
Weak reactions, on the other hand, 
operate much more slowly and thus continue to affect the nuclear composition, 
the neutrino emission, and ultimately the dynamics of the entire event. 

As compared to other 
semi-leptonic weak interactions, electron capture has a particularly 
remarkable impact on the core-collapse environment~\citep{Langanke2014305}. 
In the final stages of a star's life, the nuclear-energy generation rate 
of the core that normally sustains a star against gravitational collapse 
is absent because the core is composed of highly stable iron peak nuclei. 
Instead, at these late times the 
electron-degeneracy pressure provides the primary 
stability against collapse. It is therefore 
apparent that electron captures that remove electrons from the system 
will have dramatic consequences for this environment. 
Furthermore, the electron chemical 
potential $\mu_e$ is sufficiently large to overcome 
Q-value restrictions, and so the electron-capture rates are 
significant. 

Just prior to and during the early moments of collapse, other weak interactions 
can also play a role. \citet{MartinezPinedo2000} have shown that $\beta^-$-decay 
can temporarily compete with electron capture when $Y_e$=0.42-0.46, which can 
occur during Si shell burning and the early stages of collapse. However, as 
collapse ensues, $\mu_e$ quickly becomes large enough that $\beta$-decay electrons 
are energetically blocked due to degeneracy. Similarly, $\beta^+$-decay 
can also compete with electron-capture for nuclei with 
$Q_{e+}>2m_ec^2$, but in the core-collapse environment neutron-rich 
conditions are favored, and Q$_{e+}$ is below this threshold. 

The importance of reactions mediated by the weak nuclear force, 
and specifically electron capture, as 
it pertains to core-collapse was first demonstrated by~\citet{Bethe1979}. 
Not long after, the theory of stellar electron capture was formalized 
by Fuller, Fowler, and Newman (FFN)~\citep{Fuller1982}. 
In their pioneering work they published the first tabulation of weak 
interaction rates ($\beta^\pm$-decay and $e^\pm$-capture) considering 
presupernova conditions where allowed Fermi and Gamow-Teller (GT) transitions
dominate. Since then, advancements in computational resources have allowed 
for detailed nuclear shell-model calculations that have increased 
the accuracy of the weak-interaction theory first outlined by FFN. 
Major weak-interaction rate tabulations that derive from a combination 
of experimental data and shell-model effective interactions are the 
\citet{Oda1994} and \citet{Langanke2000} tabulations for $sd$- (A=17-39) 
and $pf$-shell (A=45-65) nuclei respectively. 
For heavier nuclei, the Shell Model Monte Carlo (SMMC) 
approach has been employed to preserve nuclear properties 
such as the correlation energy scale in very large 
model spaces~\citep{Langanke2003}. \citet{Langanke2003} 
have combined this method with an RPA technique to estimate 
electron-capture rates at densities and temperatures relevant 
during core-collapse for nuclei in the $pfg/sdg$-shell (A=65-112), 
which have come to be known as the LMSH rates. \citet{Juodagalvis2010} 
have also produced a set of more than 2200 additional rates based 
on the same RPA technique but utilizing a Fermi--Dirac parameterization 
instead of the more computationally expensive SMMC calculations. 
The individual rates were not released, but instead these rates were 
averaged over NSE abundances and reported along a characteristic 
core-collapse ($\rho,T,Y_e$) trajectory. Unfortunately, this is not suitable for 
the present study in which we investigate the detailed sensitivity 
on a nucleus by nucleus basis.

In this work, we have implemented each of the rate tabulations listed in 
Table~\ref{table:ranges}, which together 
contain 445 rates for 304 unique nuclei over a large density and 
temperature grid. This library has been built as a standalone module 
and has also been implemented into the neutrino-interaction library \texttt{NuLib} 
\citep{OConnor2015} for use in neutrino-transport routines employed by the spherically-symmetric, 
general-relativistic stellar collapse code \texttt{GR1D}~\citep{OConnor2015}--see Section~\ref{codes} for more information. Details on the density and temperature range for each of the 
included rate tabulations are shown in Table~\ref{table:ranges}. The mass coverage of each rate 
table is shown in Figure~\ref{fig:tables}.

\begin{table*}[t]
\centering
\caption{Density, temperature and mass ranges for the compiled weak rate set}
\label{table:ranges}
\begin{tabular}{*{9}{c}}
 \toprule
 \toprule
 & \multicolumn{5}{c}{Model space} \\
 \midrule
 Table & $s$ & $p$ & $sd$ & $pf$ & $pfg$/$sdg$ & T (GK) & $\mathrm{Log}_{10}(\rho Y_e$ g\,cm$^{-3})$ & Ref. \\
 \midrule
 \centering 
 FFN & x &  & x & x &  & 0.01 - 100 & 1.0 - 11 & \citet{Fuller1982} \\
 ODA & x &  & x &   &  & 0.01 - 30 & 1.0 - 11 & \citet{Oda1994} \\
 LMP & x &  &  & x &  & 0.01 - 100 & 1.0 - 11 & \citet{Langanke2003} \\
 LMSH &  &  &  &  & x & 8.12 - 39.1 & 9.22 - 12.4 & \citet{Hix2003,Langanke2001a} \\
 Approx. & x & x & x & x & x & - & - & \citet{Langanke2003} \\
 \bottomrule
\end{tabular}
\end{table*}

\begin{figure*}[ht]
 \centering
 \includegraphics[scale=1.0]{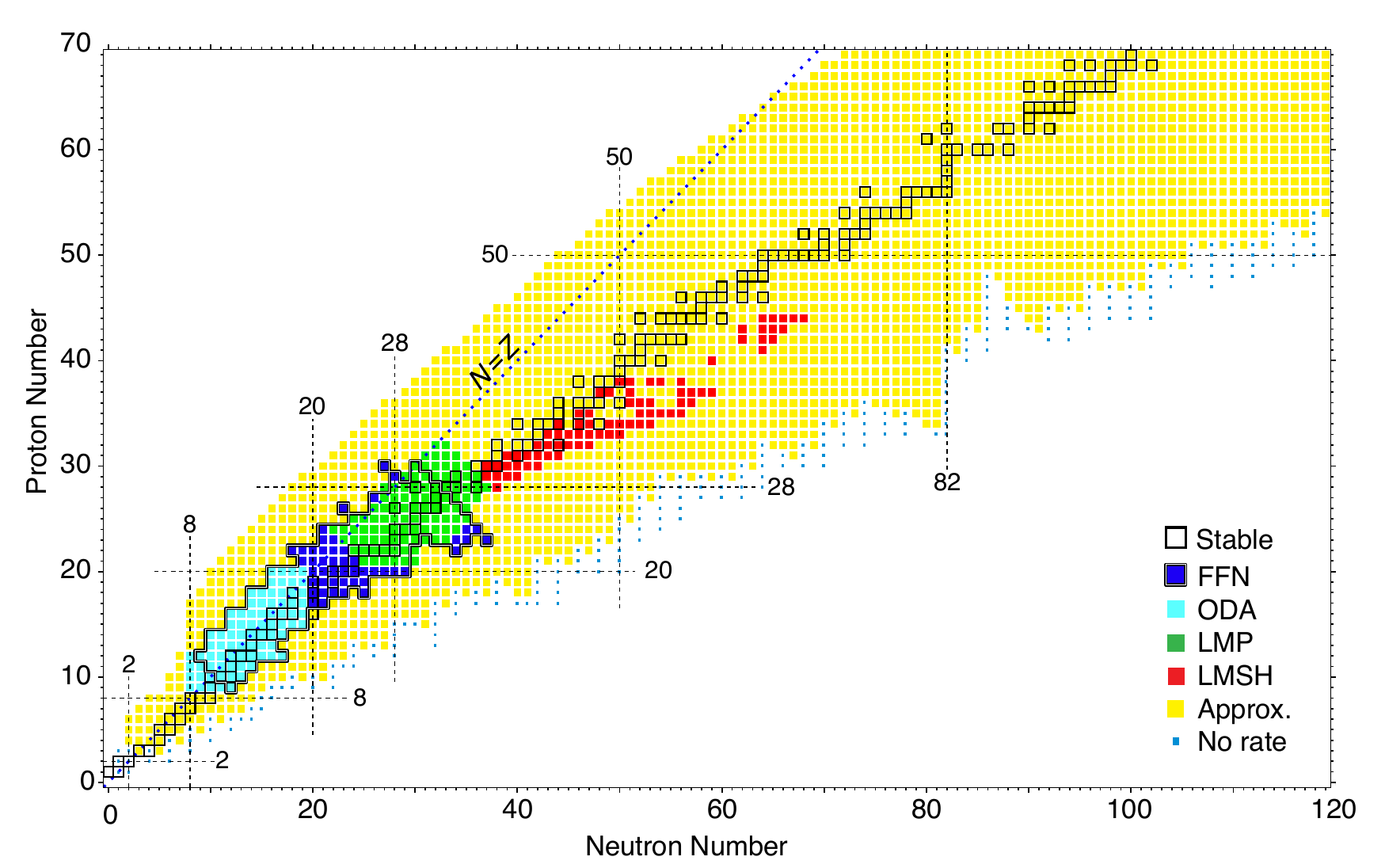}
 \caption{ Chart of the nuclear species included in each weak rate table. 
 The table to which a species belongs is given by the color and legend in
 the figure. The Oda set contains rates for lower mass $sd$-shell nuclei
 (light blue), the LMP set contains rates for the intermediate mass $pf$-shell
 nuclei (green), and the LMSH set contains rates for the heavier mass $pfg$/$sdg$-shell
 nuclei near stability (red). The FFN tabulation provides rates across the $sd$ and
 $pf$-shells (dark blue). Squares individually bordered in black are stable nuclei.
 The tables are mutually exclusive except for FFN which spans many nuclear
 shells. To distinguish between nuclei with rates from FFN and another table,
 the border of the FFN set has been outlined with a black and white line.
 }
 \label{fig:tables}
\end{figure*}

The LMP+LMSH rates were first implemented into a spherically-symmetric 
core-collapse simulation by \citet{Hix2003}. 
They compared simulations with this set of shell-model based electron-capture 
rates against simulations that utilized 
the \citet{Bruenn1985} prescription for electron capture. 
The evolution of the core-collapse phase and the structural differences 
in the core at bounce seen in that work were significant. 
In light of the differences that exist between 
theoretical estimates for electron-capture rates and those inferred 
from CE experiments, these results motivate the need for a detailed sensitivity study.

To handle the large number of nuclei not included in the tables, 
\citet{Hix2003} utilized an average electron-capture neutrino emissivity 
for all nuclei which lacked a shell-model based rate. Here, instead of 
performing averaging, we employ the approximate routine of \citet{Langanke2003}, 
which is based on the parameterization of the electron-capture rate as a function of the 
ground state to ground state Q-value. This approximation was first described by 
\citet{Fuller1985} and was later parameterized and fit to shell-model calculations 
in the $pf$-shell by \citet{Langanke2003}. In this approximation, the 
electron-capture rate is written as:

\begin{align}
\label{eq:ec}
\lambda_{\mathrm{EC}} = \dfrac{\textrm{ln}2\cdot B}{K}\left(\dfrac{T}{m_e c^2}\right)^5[F_4(\eta) 
             - 2\chi F_3(\eta)+\chi^2F_2(\eta)]
\end{align}
and the neutrino-energy loss rate is,
\begin{align}
\label{eq:nu}
\lambda_{\nu_e} = \dfrac{\textrm{ln}2\cdot B}{K}\left(\dfrac{T}{m_e c^2}\right)^6[F_5(\eta) 
             - 2\chi F_4(\eta)+\chi^2F_3(\eta)],
\end{align}
where $m_e$ is the electron mass, $K$ = 6146 s, $F_k$ are Fermi integrals of rank $k$ 
and degeneracy $\eta$, 
$\chi = (Q-\Delta E)/T$, $\eta = \chi + \mu_e/T$, and 
$T$ and $\mu_e$ are the temperature and electron chemical potential. 
$B$ (= 4.6) and $\Delta E$ (= 2.5 MeV) are fit parameters taken from \citet{Langanke2003} 
and respectively represent effective values for the transition strength and 
energy difference between final and initial excited states. 

In Figure~\ref{fig:approx}
we compare the rate estimates from this approximation and those from the tables. 
As is easily seen from the figure, the variance of the shell-model rates 
depends sensitively on the density of the environment.
At lower densities, where the electron chemical-potential and 
electron capture Q-value are comparable ($\mu_e \approx Q_{\mathrm{EC}}$), 
the location of excited states in the daughter nucleus, and the associated 
Gamow-Teller transition strength, sensitively determine the total electron-capture 
rate for a nucleus. New final states in the daughter nucleus 
become accessible as the electron Fermi-energy, which scales with $\mu_e$, 
increases beyond the energy 
required to populate them via allowed electron capture. 
The large scatter of the electron-capture rates at lower densities (Fig.~\ref{fig:approx}) 
is because the Fermi energy is comparable to the excited state energies of the 
daughter nuclei and the internal structure of each nucleus varies significantly. 
During collapse,
as the density increases and the material becomes more neutron rich due to 
successive electron captures, both the magnitude of the average 
electron capture Q-value and $\mu_e$ increase. However, $\mu_e$ 
increases more quickly with density than the reaction Q-values do, 
and eventually $\mu_e \gg Q_{\mathrm{EC}}$ implying that the majority 
of the electron-capture channels are open. 
In this regime the rate is less sensitive to the excitation energy 
spectrum of the daughter nucleus, and instead depends more strongly on the total GT strength 
across all possible final states. The decrease in the variance of the shell-model electron-capture 
rates in the higher density case of Figure \ref{fig:approx} is a result of this.

\begin{figure}
 \centering
 \includegraphics[scale=.9]{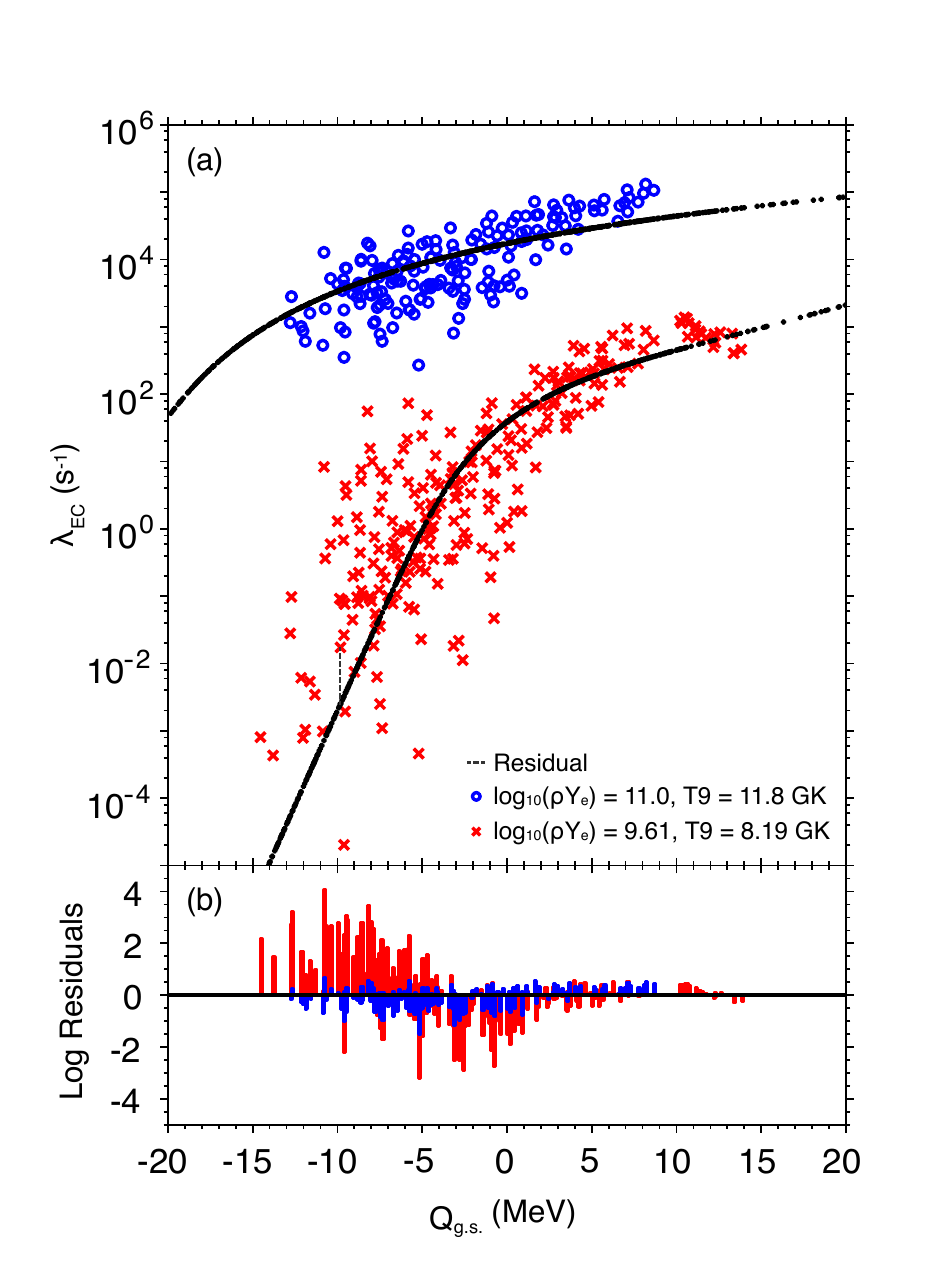}
 \caption{Panel (a): Q-value dependence of electron-capture rates at two 
 points along a core-collapse trajectory. The scattered points
 are tabulated (shell-model and SMMC) rates for each electron-capture 
 reaction, while the black points are the approximate rates given by Eq.~\ref{eq:ec}. 
 Panel (b): The residual differences between log$_{10}$ of the shell-model rates and the 
 approximate rates for each nucleus in the weak-rate library. An example residual is indicated 
 on panel (a). When the density and temperature 
 of a simulation evolve outside the range of the rate tables 
 (see Table~\ref{table:ranges}), rates are calculated via the approximate
 routines in order to avoid an artificial cut off imposed by the table boundaries.
 Rates are estimated between density and temperature grid points
 via monotonic cubic-spline interpolation as described by~\citet{Steffen1990}.}
 \label{fig:approx}
\end{figure}

While the parameters of Eqs.~\ref{eq:ec} and~\ref{eq:nu} were originally fit from the 
LMP nuclei, there is reasonable agreement of the approximation with the other tabulated rates. 
Outside of these tables, significant deviations from the estimates of this approximation 
may exist, specifically for heavier neutron rich nuclei~\citep{Juodagalvis2008}.
But for the purpose of a sensitivity study, 
this approximation---from which the majority of the rates are 
calculated---is used as a base estimate off which the 
electron-capture rates may be varied. 
In Section~\ref{species dependent} we show, given this set of rates, 
that heavier neutron-rich nuclei are the most 
important in the simulations. 
However, electron-capture rates developed from sophisticated theoretical models do not exist 
for individual nuclei in this region, and thus cannot be benchmarked against experimental measurements. 
Therefore, the estimates provided by the approximation of Eqs.~\ref{eq:ec} and~\ref{eq:nu} may 
be systematically off by a considerable amount. As we will show, 
changes in the predicted rates for these nuclei have significant consequences for the simulations, 
motivating the need for experimental and theoretical efforts to constrain the rates of 
these species.

\section{Core collapse and the role of electron capture}
\label{core-collapse}
Just prior to collapse, the temperature of the stellar core 
becomes high enough ($T\gtrsim 0.5 \, \textrm{MeV}$) that the photon gas 
has sufficient energy to photodissociate nuclei into alpha particles and free nucleons. 
However, the density is also high ($\rho\gtrsim 10^9$g\,cm$^{-3}$) resulting 
in large nuclear reaction rates that rapidly form nuclei from these light particles. 
The balance reached between these competing processes is 
known as Nuclear Statistical Equilibrium (NSE). 
If the entropy is sufficiently low and the mass fraction of free nucleons is small 
compared to that of nuclei, the most abundant nucleus in NSE is the species 
with the highest binding energy for a given electron-fraction, $Y_e$(= Z/A)
\citep{MartinezPinedo2000}. A broad distribution of abundant nuclei 
forms due to finite temperatures which distribute the abundances around these peak nuclei.

As collapse ensues and the central density increases through the first 
few decades, electron captures 
are the primary engine of deleptonization. Electrons are removed 
from the system and the produced electron-neutrinos ($\nu_e$) 
are able to freely stream out of the core, decreasing both $Y_e$ and total lepton 
fraction $Y_l$. As $Y_e$ decreases, peak abundances move toward 
neutron-rich nuclei, and the core begins to cool as $\nu_e$'s carry away energy and 
entropy. Electron captures continue to dominate the neutrino 
transport during collapse until the last few milliseconds before core bounce.
In these final moments, the central density reaches a few times 10$^{12}$ g cm$^{-3}$,  
which is large enough that the neutrino mean free path begins to shorten due 
to coherent scattering on heavy nuclei. This increase in the $\nu_e$-scattering 
cross section results in a neutrino diffusion time that exceeds the collapse time, 
thereby trapping the electron neutrinos in the inward flow of matter.
After this occurs, the conversion of electrons 
into electron-neutrinos via electron captures no longer removes 
leptons from the core. Instead, further electron captures increase 
the electron-neutrino fraction $Y_{\nu_e}$ in order to conserve the 
now constant lepton fraction and bring the system of electrons and electron 
neutrinos into equilibrium.

Prior to the work of~\citet{Langanke2001a} it was believed that electron 
captures on free protons were of greater importance than captures on nuclei 
during collapse. The main considerations involved were that electron 
capture on free protons has a higher rate owing to a smaller Q-value, 
and that nuclei with neutron number N $\geq$ 40 have full $pf$-shell 
single particle states. Thus, the addition of another neutron 
via an allowed electron-capture transition would be Fermi-blocked. 
\citet{Langanke2001a} recognized that the 
many-body nuclear states have mixed configurations and do not follow 
a simple Hartree-Fock filling of single particle orbitals. They also 
suggested that thermal excitation of nucleons to the $g_{9/2}$ orbital 
creates vacancies in the $pf$-shell, and together with 
configuration mixing, electron capture on bound protons is unblocked. 
Furthermore, because of the low entropy in the core, and the neutron-rich conditions, 
the abundance of heavy nuclei is several orders of magnitude higher than that of 
free protons, resulting in a higher overall electron-capture rate.
Thus, because electron captures on nuclei dominate, a detailed 
investigation of the contribution each species has to the deleptonization 
of the collapsing 
core, and in particular a determination of which nuclei are most important and 
therefore deserve further experimental and 
theoretical focus, would be of great value. 

\section{Codes \& Methods}
\label{codes}
\subsection{\texttt{NuLib}} 
In addition to the electron captures rates developed in this paper,
other rates are needed to perform core-collapse simulations.  The
collection of rates we use is contained in \texttt{NuLib} \citep{OConnor2015},
an open-source, neutrino-interaction library available as a GitHub
repository at http://www.\texttt{NuLib}.org. \texttt{NuLib} contains routines for
calculating electron-type neutrino/antineutrino charged-current
absorption opacities on nucleons with corrections for weak magnetism and 
nucleon recoil based on the formalism of \citet{Burrows2006} and
\citet{Horowitz2002}. Neutrino emissivities for these processes are
determined via Kirchhoff's law which equates the absorption rate of a
equilibrium neutrino distribution to the emission rate of the
underlying matter. Elastic scattering of neutrinos on nucleons, and
coherent scattering of neutrons on alpha particles and heavy nuclei is
also included in \texttt{NuLib}. For the former we include corrections 
for weak magnetism and nucleon recoil, and for the latter we include corrections
from ion-ion correlations \citep{Horowitz1997}, electron polarization,
and the nuclear form factor. Inelastic scattering of neutrinos on
electrons is included based on the expressions of
\citet{Bruenn1985}. Emissivities of heavy-lepton neutrino/antineutrino
pairs via electron-positron annihilation and nucleon-nucleon
Bremsstrahlung are computed ignoring final state neutrino
blocking. For neutrino-antineutrino annihilation, instead of computing
the non-linear absorption opacity during the simulation, we make use
of an effective absorption opacity, which has been shown to be an
excellent approximation for core-collapse supernovae
\citep{OConnor2015}.

\subsection{\texttt{GR1D}}
We test our electron-capture rate implementation, and study the
sensitivities of the core-collapse phase to these rates using the code \texttt{GR1D}
\citep{OConnor2010,OConnor2015}. \texttt{GR1D} is an open-source
spherically-symmetric general-relativistic neutrino-transport and
hydrodynamics code used for studying stellar collapse and the early
stages of a core-collapse supernova. For details of the hydrodynamics
module of \texttt{GR1D} we refer the reader to \cite{OConnor2010}. The neutrino
transport is handled though a general-relativistic, energy-dependent
two-moment formalism for which extensive details can be found in
\citet{OConnor2015}. Our scheme numerically solves for the time
evolution of the first two moments of the neutrino distribution
function: the neutrino energy density and the neutrino momentum
density. The simulations utilize 18 energy groups
logarithmically spaced between 0 and 250\,MeV. 
Only electron type neutrinos are evolved until the central density reaches
$10^{12}$\,g\,cm$^{-3}$, after which electron anti-neutrinos and a
characteristic heavy lepton neutrino are included. However, these latter two
neutrinos do not become important until core bounce has occurred.
Spatial fluxes of the neutrino moments are treated explicitly.  Inelastic neutrino-electron
scattering is handled explicitly until the central density reaches
10$^{12}$\,g\,cm$^{-3}$ at which point an implicit treatment is used. 
Together \texttt{NuLib} and \texttt{GR1D} provide a robust and extendable 
code base, making them ideal for the present study.

\subsection{Neutrino emission via electron capture}
Electron capture is associated with the emission of electron neutrinos 
and so the electron-capture rate is proportional to the integrated 
spectrum of $\nu_e$ emitted per second.
The rate for a particular nuclide, as tabulated in the implemented rate tables, 
is defined as the sum of the rates for each of the individual nuclear transitions 
\begin{align}
\lambda = \sum\limits_{ij} \lambda_{ij},
\end{align}
where indices $i$ and $j$ correspond to levels in the parent and daughter 
nucleus respectively. The spectra of emitted neutrinos from the 
electrons capturing on nuclei, described by the matter temperature $T$ and 
electron chemical potential $\mu_e$, will vary based on the initial and final 
states involved owing to a different reaction Q-value,
\begin{align}
Q^{\mathrm{EC}}_{ij} = Q_{g.s.} + E_i - E_j
\end{align}
where $Q_{g.s.}$ is the atomic mass difference of the initial and 
final nuclei, and $E_i$ and $E_j$ are the excitation energies of 
the populated states in the parent and daughter nucleus respectively.
The most comprehensive solution to constructing neutrino spectra 
would be to coherently sum the spectra of neutrinos emitted from each 
nuclear transition. However this would rely upon rate tabulations 
for individual transitions which are not presently available. 
Thus, we implement an effective neutrino spectra 
in terms of a single reaction Q-value, $q$, that is chosen to constrain 
the average energy of the spectrum to match that from the tabulated 
rates~\citep{Langanke2001b}, 
\begin{align}
n(E_\nu,q)=E_\nu^2(E_\nu-q)^2\dfrac{N}{1+\textrm{exp}\{(E_\nu-q-\mu_e)/kT\}}
\end{align}
\begin{align}
\label{eq:avge}
\langle E_\nu \rangle = 
\dfrac{\int\limits_{0}^{\infty}E_\nu n(E_\nu,q) dE_\nu}{\int\limits_{0}^{\infty}n(E_\nu,q) dE_\nu} =
\dfrac{\lambda_\nu}{\lambda_{\mathrm{EC}}+\lambda_{\beta^+}},
\end{align}
where $n(E_\nu,q)$ is the neutrino distribution function 
and is normalized to the total electron-capture rate for a particular nuclear species. 
$\lambda_\nu$, $\lambda_{\beta^+}$, and $\lambda_{\mathrm{EC}}$ are the neutrino energy loss, 
positron emission, and electron-capture rates respectively. 
Eq.~\ref{eq:avge} is solved numerically for the effective Q-value, $q = q_{\textrm{eff}}$, 
which then defines the effective neutrino spectrum for the electron-capture 
reaction of interest at a given $\rho$, $T$, and $Y_e$. The approximate 
neutrino spectra generated in this way are 
unable to reproduce complex structure such as double peaking in the true neutrino 
distribution, which may occur when there is a resonant 
allowed transition ($Q^{\textrm{EC}}_{ij}\sim 0$) between an excited parent 
state and the daughter-nucleus ground state. However, it approximates 
singly-peaked neutrino distributions quite well~\citep{Langanke2001b}. The 
spectrum is normalized to the total electron-capture rate via Gaussian-Legendre 
quadrature with an algorithm we developed to adaptively adjust the range of integration 
to the full width of the spectrum.

\begin{figure}
 \centering
 \includegraphics[scale=1.0]{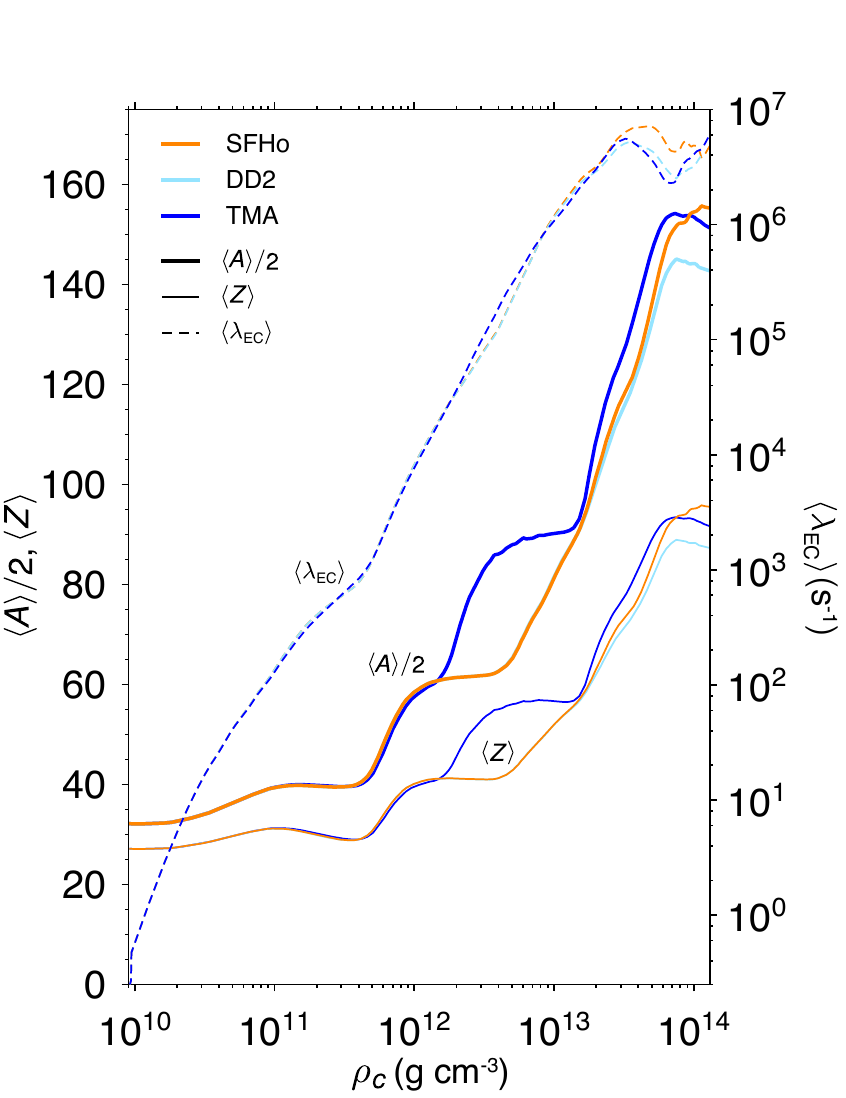}
 \caption{ The average nuclear mass (divided by two),
 charge, and electron-capture rate versus central density
 for the three EOS utilized in this study. The colors indicate 
 different EOS, while the line style indicate which quantity is 
 plotted. All three EOS
 have nearly identical abundance distributions up to densities
 of 2$\cdot$10$^{12}$ g cm$^{-3}$. Beyond this point the TMA EOS
 has a heavier and slightly more neutron rich
 mass distribution compared to both SFHo and DD2, but maintains
 a comparable average electron capture rate overall. These simulations
 each utilize the s15WW95 progenitor.}
 \label{fig:eos_mass}
\end{figure}

Utilizing these spectra, the electron-capture neutrino emissivity for a 
given nuclear species is calculated as
\begin{align}
\eta_i(E_k) = \dfrac{1}{4\pi} E_k n_i n(E_k,q_{\textrm{eff}}),
\end{align}
where the $\nu_e$'s are assumed to be emitted isotropically, $n_i$ is the 
number density for the $i$-th nucleus, $n(E_k,q_{\textrm{eff}})$ is the neutrino 
spectra evaluated at the effective Q-value that solves  Eq.~\ref{eq:avge}, 
and $E_k$ indicates the energy of energy group $k$. 
Evaluation of the emissivity is done point wise at the centroid of each energy bin,
and has units erg/(cm$^3\cdot$sr$\cdot$s$\cdot$MeV). 

For estimates of the NSE number densities used above, we 
utilize several EOS from~\citet{Hempel2010}. In particular, we use the SFHo 
EOS and internally consistent NSE 
distribution developed by~\citet{Steiner2013} for our primary 
EOS. We also compare with the DD2~\citep{Typel2010} 
and TMA~\citep{Toki1995} EOS, each with self-consistent, but different, 
NSEs. The SFHo and DD2 EOS were chosen because they currently best 
satisfy both nuclear and astrophysical constraints~\citep{Fischer2014}. 
Instead of meson self-interactions, the DD2 EOS implements 
density-dependent meson-nucleon couplings which have been 
used successfully to describe nuclear structure in a wide region of the nuclear 
chart and have also been tested in heavy-ion collisions~\citep{Typel2010}.
For nuclear masses, the SFHo and DD2 NSE distributions rely on the 
Finite Range Drop Model (FRDM) from \citet{Moller1995} and \citet{Moller1997}, 
whereas the TMA EOS utilizes a mass table calculated by~\citet{Geng2005}.
For consistency, in addition to NSE abundances, these mass distributions 
are also utilized in the calculation of reaction Q-values for 
use in Eq.~\ref{eq:ec}.

\section{Sensitivity study} 
\label{sensitivity}
\subsection{Reference simulations}
In order to establish reference simulations off which variations are performed, 
we have utilized the widely studied 15 solar mass, solar metallicity 
progenitor s15WW95~\citep{Woosley1995}, as well as s12, s20, and s40 from WH07~\citep{WOOSLEY2007}, 
which span the range of stellar compactness $\xi_{2.5}$\citep{OConnor2011} in this model set. 
For more details, see the progenitor sensitivity 
subsection below. For each progenitor we use the SFHo EOS described above, 
and in addition, for s15WW95 we also employ the DD2 and TMA EOS and 
NSE distributions. 

For each reference simulation, a 
full complement of neutrino-interaction microphysics is incorporated via \texttt{NuLib}, 
which includes the newly implemented weak rates library described here. 
The weak-rate tables were included using the following priority hierarchy: 
$LMP > LMSH > Oda > Approx.$, ensuring that rates from sources with higher priority are
utilized where rate estimates from multiple sources exist. $Approx.$ indicates the parameterized rate 
approximation of Eqs.~\ref{eq:ec} and~\ref{eq:nu}, which is used for nuclei not included in the tables 
and for regions of density and temperature which are beyond the 
limits found in Table~\ref{table:ranges}. For consistency, only tables that 
derive from shell-model calculations are utilized.

For each progenitor and EOS we perform collapse simulations in {\tt{GR1D}} and follow the 
evolution until at least $\sim100\,$ms after bounce. The collapse proceeds as described 
in Section \ref{core-collapse}. Differences in the collapse evolution for different progenitors 
stem from the hydrostatic conditions in the cores of these massive stars at the onset of 
collapse. For stars with large $\xi_{2.5}$, larger central temperatures are needed to 
balance gravity. This gives lower central densities, and therefore less electron capture 
during the final stages of stellar evolution. The range of initial central $Y_e$ goes from $\sim$0.422 for 
the s15WW95 model to $\sim$0.447 for the s40WH07 model, or a range of $\sim$6\%. 
After neutrino trapping sets in, we see a range of trapped lepton-fraction of 
$\sim$0.288\,--\,0.297, where s40WH07 and s12WH07 have the minimum and maximum trapped 
$Y_l$, respectively. The overall higher deleptonization rate for the more compact 
progenitors is due to both longer collapse times and larger matter temperatures, which 
enhance the electron-capture rates.

Simulations utilizing different EOS, while holding all else constant, demonstrate 
only small variations in the density, temperature, and $Y_e$ central-zone trajectories 
up to bounce. Figure~\ref{fig:eos_mass} details the abundance distributions for each EOS, 
as well as the resulting average electron capture rate along a collapse trajectory. 
The NSE distributions of all three EOS are largely similar early on, but 
differences in the mass table of the TMA EOS cause it to diverge from the others starting around 
$1-2\times10^{12}$g\,cm$^{-3}$. However, differences are seen in the electron-capture 
rate only after central densities of $2\times10^{12}$g\,cm$^{-3}$, 
where any effect on the evolution is suppressed because of neutrino trapping.
Near nuclear saturation density, however, the differences in EOS 
begin to play a more important role. The density-dependent couplings of the DD2 EOS, for 
instance, result in higher central temperatures at bounce. However, since the average rate in simulations 
utilizing each of the EOS are nearly identical, they result in 
a difference of trapped lepton-fraction of only a fraction of a percent. For more 
information on the sensitivity of the electron and lepton fractions to the 
EOS during collapse see \citet{Fischer2014}. 

Together, these reference calculations span a wide range of progenitor and EOS dependences 
that ensure a configuration-independent assessment of the core-collapse sensitivity 
to electron capture on nuclei, and furthermore demonstrate the universality of collapse. 
In what follows, we detail the results primarily for 
variations on the s15WW95+SFHo reference simulation, but describe any significant differences, where they 
exist, in relation to variations on the other progenitor+EOS reference simulations.

\subsection{Species dependent sensitivity}
\label{species dependent}
To understand the sensitivity of core-collapse to different 
regions of electron capturing nuclei, we use the central zone collapse profile 
from our reference simulation 
to decompose the change of the electron fraction with time, $\dot Y_e$, into 
the electron captures of each nuclear species. While using only the central zone 
is an approximation, we justify this by noting the observation by \citet{Liebendorfer2005} 
that the electron fraction profiles typically correlate quite well with density during the 
collapse phase.  Therefore, matter will generally have the same electron capture history.
We calculate $\dot Y_e$ and account for $\nu_e$ re-absorption in the following energy-dependent way,
\begin{align} \label{eq:yedot}
\dot Y_e^i = \frac{4\pi\alpha}{\rho N_a}\sum\limits_k \frac{\Delta \epsilon_k\cdot\eta_i(\epsilon_k)}{\epsilon_k}\cdot \left(1-\frac{E_k}{B_k}\right)
\end{align}
where $\dot Y_e^i$ is the time derivative of the electron fraction due to electron captures on the $i$th 
nuclear species, $\alpha$ accounts for the general relativistic time dilation, $N_a$ is Avogadro's constant, 
$\rho$ is the density, $\epsilon_k$ is the energy of the $k$th energy bin, $\Delta \epsilon_k$ is the $k$th energy 
bin width, $\eta_i$ is the emissivity of species $i$ and $1-\frac{E_k}{B_k}$ is the neutrino 
blocking factor that accounts for re-absorption as
collapse approaches weak equilibrium. Along with a hydrodynamical correction due to advection 
of electrons into the central zone, the time integral of Eq.~\ref{eq:yedot} added 
for all nuclei reproduces the full time dependent $Y_e$ profile of the central zone during 
collapse, indicating that electron captures on heavy nuclei singularly drive the deleptonization 
of the central zone.

With this method, the deleptonization 
history due to each nucleus can be individually investigated. 
At these densities and temperatures, NSE diversifies 
the abundant nuclei, ensuring that no single nucleus dominates the deleptonization. 
There are, however, subsets of nuclei that contribute more than others 
to the reduction of $Y_e$. A nuclear-mass dependence can be studied by binning the 
the contribution to $\vert \dot Y_e \vert$ from each nuclide into nuclear mass bins 
and tracking the evolution of each region up to neutrino trapping. 
Figure~\ref{fig:yedot}a plots the deleptonization rate in the 
core for different nuclear mass bins, as the central $Y_e$ progresses from its 
progenitor value to its value when weak equilibrium is achieved, just prior to bounce. 
Early on, before the collapse becomes strongly dynamical, nuclei in both the mass 
range 25\textless A\textless 65 ($sd$+$pf$-shell) and those in the 
65\textless A\textless 105 ($pfg$/$sdg$-shell) comprise the main component of 
the deleptonization. However, during the strongest 
push toward neutron-rich conditions, where $Y_e$ rapidly changes from $\sim$0.41 to 
$\sim$0.28, nuclei with mass A \textgreater 65 dominate the evolution as seen by the 
red and light blue curves in Figure~\ref{fig:yedot}a. Unfortunately, the most precise 
electron-capture rate estimates fall below this region and instead, the rates 
are set primarily by the approximation of Eq.~\ref{eq:ec}. 

\begin{figure}
 \centering
 \includegraphics[width=1. \columnwidth]{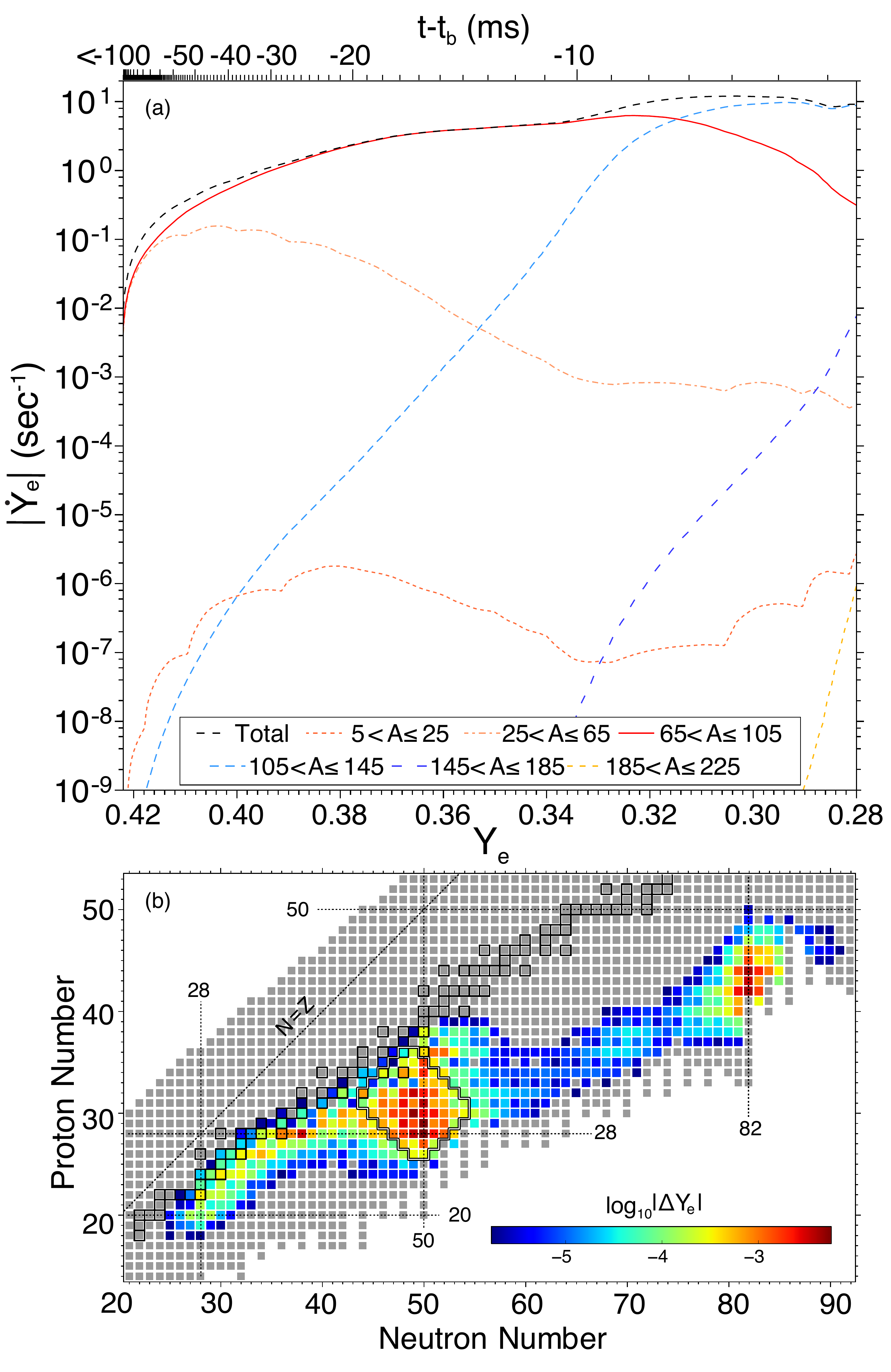}
 \caption{ (a) The contribution of nuclear electron capture to the change of the 
 matter electron-fraction with time. The contours are the binned sums of
 $\vert \dot Y_e \vert$ for each species in several mass regions.
 For reference, the central density at $t-t_b$ = -20, -10, -5, -2, and -1 ms
 is $1.41\cdot 10^{11}$, $4.06\cdot 10^{11}$, $1.42\cdot10^{12}$, $8.49\cdot 10^{12}$,
 $3.30\cdot 10^{13}$ g cm$^{-3}$ respectively. 
 (b) Top 500 electron capturing nuclei with the largest absolute change to the electron 
 fraction up to neutrino trapping. The color scale indicates
 $\vert \dot Y_e \vert$ integrated up to the trapping time, occurring when 
 $\rho_c \sim 2\cdot 10^{12}$ g cm$^{-3}$, such that the total electron-fraction 
 at this point is equal to 
 its initial value less the sum of $\Delta Y_e$, the plotted quantity, over all nuclides.
 Calculations are based on the s15WW95+SFHo reference simulation. The rectangular outline indicates 
 the size of the sampling region used in the statistical resampling study, and 
 also the set of nuclei which exhibited the largest changes to the simulations
 when excluded from the electron-capture calculations.}
 \label{fig:yedot}
\end{figure}

It is also useful to understand the specific nuclei that have the largest 
integrated contribution to core deleptonization up to neutrino trapping. 
In panel (b) of Figure~\ref{fig:yedot} we plot the 500 nuclei with 
the largest integrated $\vert \dot Y_e\vert$ from $t=0$ to the trapping 
time---when densities are in excess of $2\cdot10^{12}$ g cm$^{-3}$. This 
reveals the channel through which the bulk of electron captures 
operate. The central electron-fraction at the trapping density is reproduced 
by subtracting the sum of this quantity over all nuclear species from the initial 
electron-fraction,
\begin{align} 
Y_e(t=t_\textrm{trapping}) \simeq Y_e(t=0) - \sum\limits_i \Delta Y_e^i,
\end{align}
where $\Delta Y_e^i$ is shown in Figure~\ref{fig:yedot}b, and the component of 
$Y_e$ due to advection of electrons into the central zone (otherwise making this 
relation exact) is left out for simplicity. 
Within the $pfg$ and $sdg$-shells we find that primary contributors to the 
deleptonization phase of collapse are neutron rich nuclei near the 
N=50 and N=82 closed neutron shells.

To confirm these results, we also gauge the sensitivity of the collapse phase to localized 
groups of nuclei by employing a statistical resampling technique where sets 
of nuclei are removed from the simulation. This method 
is based on well known statistical resampling methods such as 
bootstrap and jackknife resampling \citep{Wu1986}. 
Specifically, a rectangular region centered on a nucleus and spanning all 
nuclei within $\pm 3$ isobars and $\pm 5$ isobaric chains is removed from the 
calculation of the electron-capture neutrino emissivity. An example of such a 
removed region is drawn on Figure~\ref{fig:yedot}b. This technique 
is employed in 48 simulations with resampling performed uniformly across the 
nuclear chart. Using this technique we find the simulations are most sensitive to 
nuclei in the mass range 74-84 with $Z/A$ ($=Y_e$) between 0.36-0.44, corresponding 
to nuclei near $\nuc{78}{Ni}$, $\nuc{79}{Cu}$, and $\nuc{79}{Zn}$. 
These results agree with the $\dot Y_e$ calculations performed 
above, and indicate that species near the N=50 magic number have the largest 
contribution in magnitude to the change in the electron fraction overall. 
The impact of removing these species from the simulation 
corresponded to a change of inner-core mass at bounce of $\gtrsim$ 10\%, 
whereas resampling in other regions resulted in variations 
of only a few percent.  
 
\begin{figure}
 \centering
 \includegraphics[scale=.7]{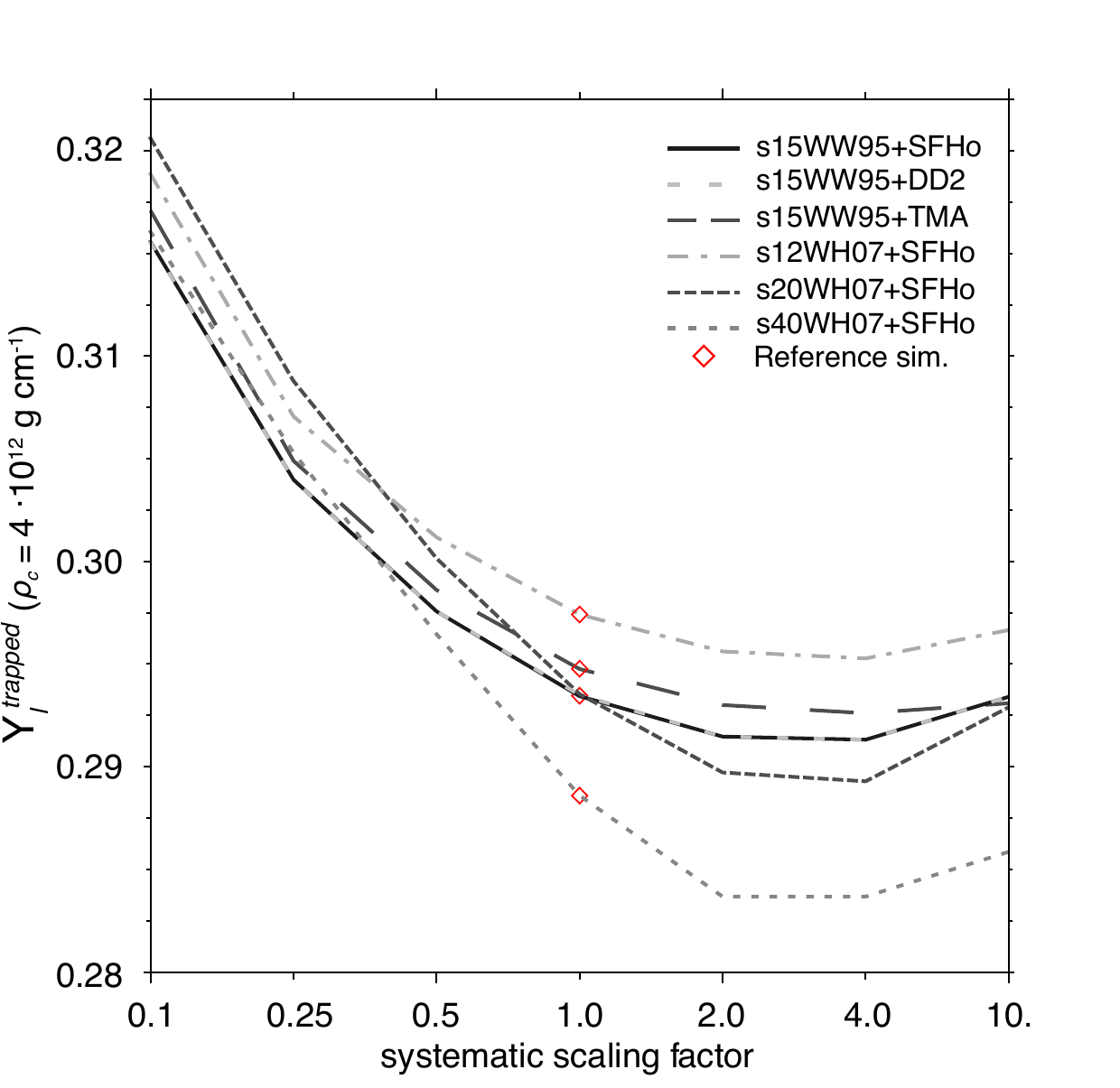}
 \caption{ Projection of the (trapped) lepton fraction at
 $\rho_c = 4\cdot 10^{12}$\, g\,cm$^{-3}$
 as a function of the electron-capture rate scaling factor 
 for progenitor+EOS reference simulation. In all
 the cases the lepton fraction begins to increase if the
 capture rate becomes too high because of a dramatic
 increase in the electron neutrino absorption cross
 section. The asymmetry seen here indicates that those
 quantities which depend on $Y_{l/e}$ are likely to be more
 sensitive to a reduction of the electron-capture rates
 due to a systematic overestimate in the base rates, than they are to
 an increase due to an underestimate.}
 \label{fig:ylepscale}
\end{figure}

The electron-capture rates for these  
nuclides rely entirely on the approximation of Eqs.~\ref{eq:ec} and~\ref{eq:nu}, which 
were fit originally to rates of lower-mass mid-shell nuclei near stability. 
Therefore, in the region indicated by the above two studies, the approximation is 
largely uncertain and may be systematically off by a significant amount. For instance, 
these estimates do not account for nuclear structure effects that may occur near the 
N=50 closed neutron shell. Depending on the nuclear configurations, 
thermal excitations, and increasing dependence on forbidden transitions, 
Pauli blocking may considerably reduce the electron-capture rates in this area. 
Given that the change of inner-core mass at bounce was largest when the 
rates of these nuclides were decreased to zero as compared to any other set, 
and that without any evaluative measurements the uncertainties in these rates remain large, 
experimental and theoretical work should focus here. Any substantial changes to the 
electron-capture rate estimates for these nuclei will likely have a relatively large 
impact on simulation predictions for the PNS formation, and will therefore help to 
constrain important collapse and pre-explosion phase quantities.

\subsection{Systematic variations}
\label{Systematic}
To study the strongest impact of variations in the electron-capture rates, we 
perform simulations in which the rate for each A\textgreater 4 nuclide is systematically 
scaled by factors of 10, 4, 2, 0.5, 0.25, and 0.1. In this way, the 
structure of the rates as seen in Figure~\ref{fig:approx} is preserved 
(the lower panel of residuals is unaffected), but the distribution of rates is 
shifted to larger or smaller values depending on the scaling factor. 
Systematic shifts of the rates emphasize the role of electron capture 
as a regulator for entropy and temperature in the simulations. By increasing the rates, 
more neutrinos are emitted and escape during the initial stages of collapse, thereby 
increasing the evaporative neutrino-cooling. 
Furthermore, because the dominant source of matter pressure is electron degeneracy, 
increased electron-capture rates accelerate the collapse. This impacts the matter 
profiles outside the shock in the early post-bounce phase.
Decreasing the rates has the opposite effect, the entropy, temperature, and electron fraction 
of the core are significantly higher because less cooling takes place.

The evolution prior to and right at $\rho_c = 2\cdot 10^{12}$\, g\,cm$^{-3}$ 
(which is the density that defines neutrino trapping) is what sets the 
final value of the trapped lepton and electron fractions, which are 
important due to their direct impact on the formation of the PNS.
For all the reference simulations we observe a minimum in the trapped 
lepton fraction occurring for a systematic scaling factor of approximately four. 
The minimum that forms can be seen in Figure~\ref{fig:ylepscale}. 
Scaling by ten slightly reverses the downward trend, and 
increases the trapped lepton-fraction from its minimum value. This behavior is the result of 
electron-neutrino capture on heavy nuclei becoming the primary source of opacity, 
exceeding what is typical as a result of coherent $\nu_e$-scattering. When the rates 
have been enhanced by a factor of ten, the ratio of the absorption and scattering 
opacities, $\kappa_a / \kappa_s$, surpasses unity already by central densities of 
$3\cdot 10^{11}$ g cm$^{-3}$ and $\kappa_a\sim 4\kappa_s$ by the time 
$\rho_c = 10^{12}$g cm$^{-3}$. Absorption cross sections 
are then large enough to trigger an early onset of neutrino trapping at densities 
lower than what is found for the reference rates. The consequence is that electron capture 
has a smaller window of deleptonization, leading ultimately 
to a higher $Y_l$ overall.

\begin{figure}
 \centering
 \includegraphics[scale=.55]{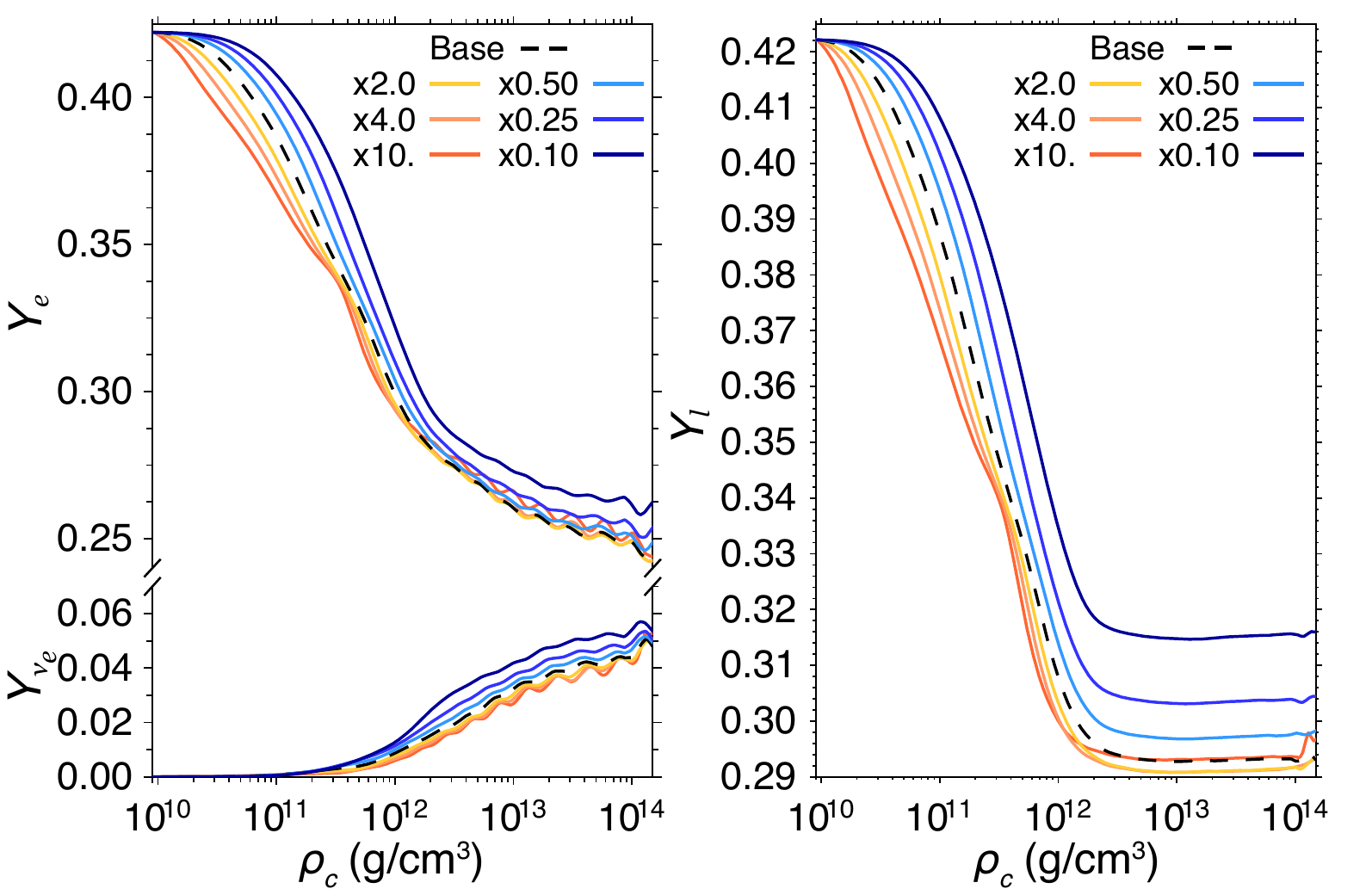}
 \caption{ Comparison of the central electron,
 electron-neutrino (left) and lepton (right)
 fractions in which the nuclear electron-capture
 rate for every species has been scaled by factors shown in
 the legend. Warmer colors indicate a higher overall electron
 capture rate, and cooler colors indicate a lower rate. The
 dashed black line indicates the reference s15WW95+SFHo simulation.}
 \label{fig:yrho}
\end{figure}

\begin{figure*}
 \centering
 \includegraphics[scale=0.9]{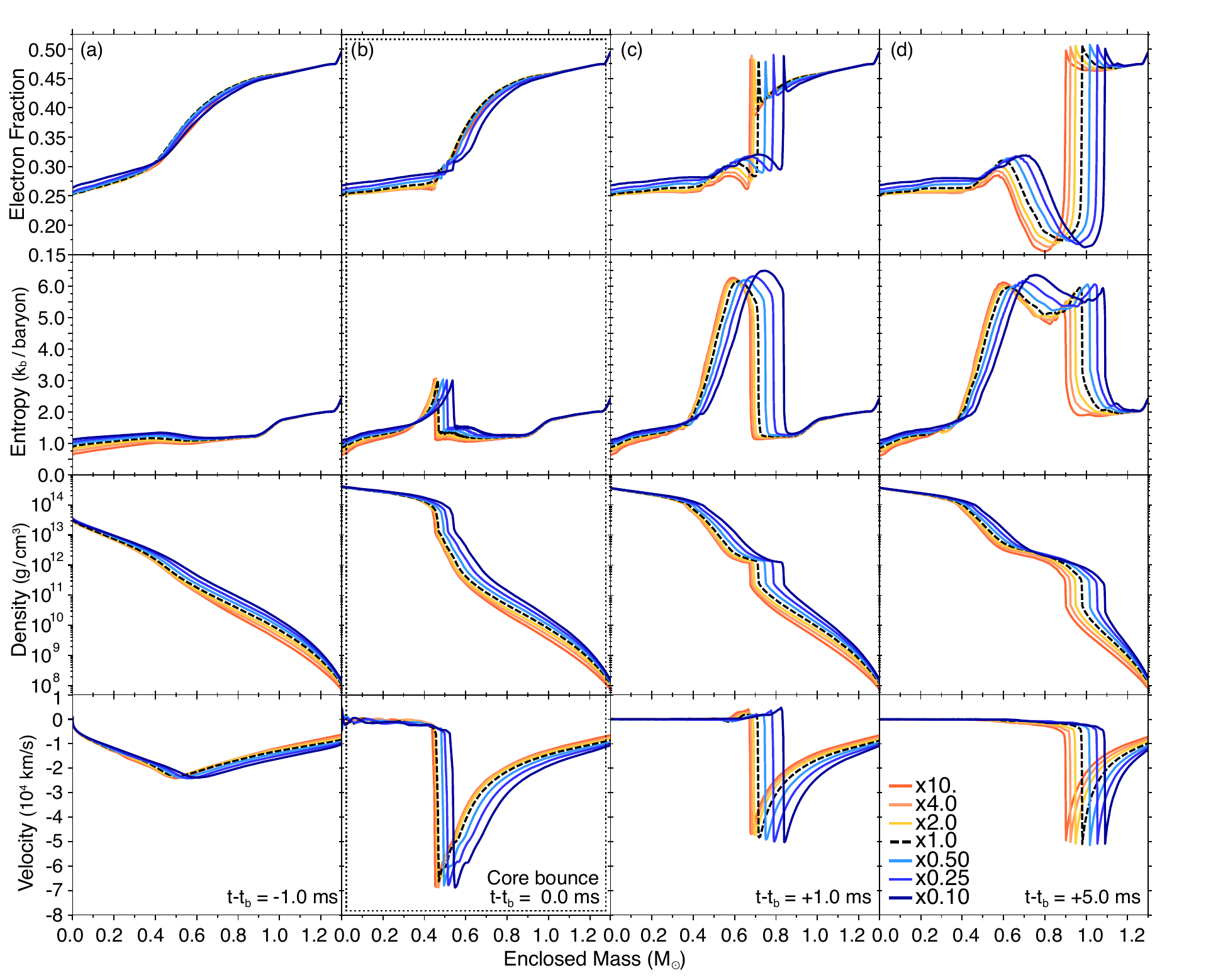}
 \caption{ The electron fraction, entropy, density, and velocity as a function of
  enclosed mass at four times during a core-collapse simulation, spanning 6 ms 
  around bounce, including the collapse phase just after the onset of neutrino trapping
  (a), core bounce (b), and 1 ms (c) and 5 ms (d) after bounce during which the shock has 
  begun its outward trajectory. The reference 
  simulation (s15WW95+SFHo) is shown in dashed black. Simulations shown in color have 
  electron-capture rates scaled systematically for all species by factors of 
  10, 4, 2, 0.5, 0.25, and 0.1. Core bounce ($t-t_b = 0$) shown in panel (b) is 
  defined as when the entropy at the shock front exceeds 3.0 k$_b$/baryon.}
 \label{fig:hydro_mass}
\end{figure*}

The range of electron fractions near core bounce is commensurate with the 
range of trapped lepton-fractions so far described, see Figure~\ref{fig:yrho}.
As mentioned above, variations of $Y_e$ (and $Y_l$) on this level are of 
importance due to its direct impact on the formation of the PNS and the supernova shock. 
Electron fraction, entropy, density and velocity profiles are shown in 
Figure~\ref{fig:hydro_mass} for s15WW95+SFHo at -1, 0, 1, and 5 milliseconds 
relative to bounce. Of particular interest, 
we find that the mass of the forming PNS inner-core at bounce, seen as the mass behind the 
steep velocity gradient in panel (b), varies on the order of $\sim$0.1 $M_\odot$, 
and up to $\sim$0.2 $M_\odot$ five milliseconds after bounce. The asymmetry observed in 
the trapped-lepton fraction, where scaling the rates by 0.1 had a more dramatic effect than 
scaling by 10, translates directly to the variation of the inner-core mass at bounce 
(+16/-4\,\% from the reference). 
The result we find is that the forming PNS has a lower bound on the inner-core mass at 
bounce over the range of electron-capture rates explored.
Because the rates are already high, and therefore the absorption 
opacity is already almost comparable to the scattering opacity, the range of 
inner-core mass at bounce comes mainly from simulations with decreased rates 
relative to the base simulation. 

In addition to the direct impact on core dynamics and structure, the neutrino emission 
at bounce is found to be very sensitive to these variations. 
Figure~\ref{fig:nulum} shows the neutrino luminosity 500 km 
from the center for the different neutrino species as a function of time. 
Prior to bounce the $\nu_e$-luminosity begins to rise from 
electron captures on bound protons in nuclei, 
but is quickly regulated by neutrino trapping, causing a down 
turn in the luminosity. During this time the core is very sensitivite to the 
nuclear electron-capture rates as the entropy is low enough that heavy nuclei dominate 
the available mass. Scaling the rates for each nucleus using the same systematic factors 
results in a 40\% variation of the $\nu_e$-luminosity before bounce.
During bounce, the electron-neutrino burst---seen as the peak luminosity in the left panel of 
Figure~\ref{fig:nulum}---is powered primarily by electron capture on free protons. The 
core-bounce and shock liberates nucleons from their bound states and the entropy rises causing 
a significant increase in the nucleon and light particle abundances. 
That said, while the electron-capture rate on free protons, $\lambda_p^{\mathrm{EC}}$, 
is not adjusted in these simulations, a range of $\pm$20\% relative to the reference peak 
$\nu_e$-luminosity is observed.

\begin{figure}
 \centering
 \includegraphics[scale=.5]{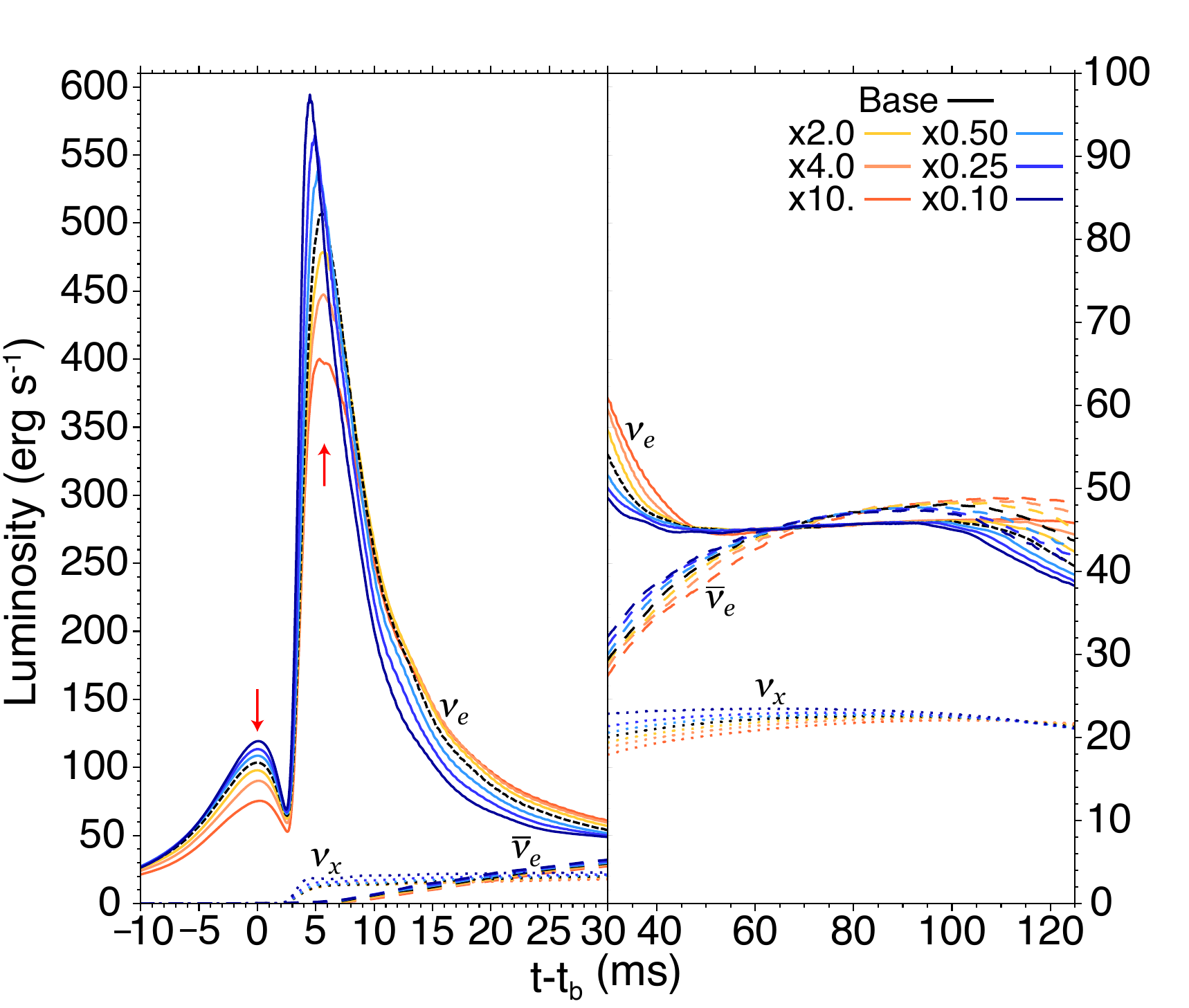}
 \caption{ The neutrino luminosity as measured at a radius of 500 km as a
  function of time after bounce in our s15WW95+SFHo
  simulation set. Electron-capture rate
  scaling factors are shown in the legend, where contours with
  warmer colors have higher rates, and cooler colors have lower
  rates. While the peak electron-neutrino
  luminosity is considered particularly stable across core-collapse
  simulations, it varies significantly with
  variations of the electron-capture rates on medium-heavy
  nuclei. When the rates are at their lowest ($\times$0.1 case), the shock reaches
  the neutrinosphere more quickly than in the other simulations.
  This results in a larger luminosity in the peak electron-neutrino burst 
  because more $\nu_e$s are able stream out of the core at early times.
  The opposite is true when the rates are higher, the neutrinosphere 
  and shock converge much more slowly, and so the neutrinos spend more 
  time diffusing out of the inner core, reducing the peak luminosity 
  but distributing it out to later times.}
 \label{fig:nulum}
\end{figure}

We find that these dramatic variations of the peak electron neutrino luminosity 
are a result of alterations to the neutrinosphere and shock convergence-timescale.
Specifically, when electron captures on nuclei are weaker (scaling by 0.1), the 
inner-core mass that forms at bounce is significantly larger. 
This results in more kinetic energy transferred to the shock, allowing it to sweep up 
mass more quickly. In Figure~\ref{fig:hydro_mass} this 
can be seen by the broadening of the distribution of shock locations in mass 
between the different 
simulations in the velocity plot 5 ms after bounce (bottom-right) as compared to $t-t_b$ = 0.
Also, with a weaker overall rate the opacity will be lower, allowing the 
neutrinosphere to move in to lower radii more quickly. The combination of these effects 
result in the shock and neutrinosphere radii converging earlier for the simulations 
with lower electron-capture rates, and later for simulations with higher rates, 
up to a difference on the order of 3.5 ms. Thus, electron capture on protons 
liberated by the shock produce neutrinos that are able to reach the neutrinosphere 
earlier and freely stream away, contributing to a larger $\nu_e$ peak luminosity 
when the nuclear electron capture rate is systematically lower. On the other hand, 
when the nuclear electron capture rate is high, the emitted neutrinos diffuse more 
slowly through the core, and reach the neutrinosphere at later times, thus 
strongly quenching the peak luminosity but spreading out the emission to later 
times. Due to the high luminosity of the electron-neutrino burst near the time of bounce, 
it is a candidate for detection from a galactic core-collapse supernovae 
in Earth-based detectors sensitive to electron neutrinos, e.g. 
those with a detector volume composed of liquid Argon. 
And while such measurements are not presently of high 
enough precision to resolve each variation seen here, they may indicate the 
total amount of electron capture occurring at core bounce.

\begin{figure}
 \centering
 \includegraphics[width=1. \columnwidth]{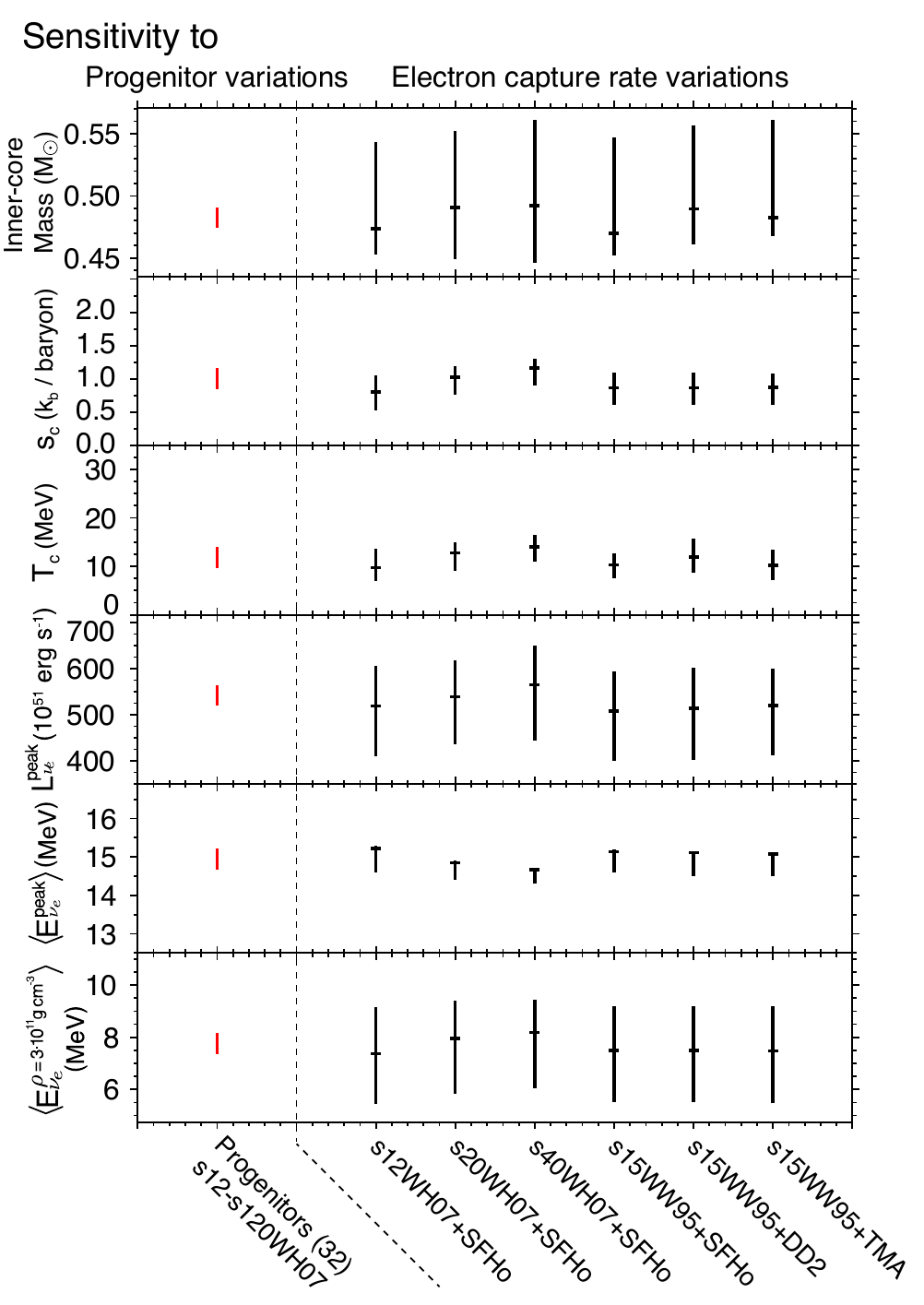}
 \caption{ The full range of sensitivity of the PNS inner-core mass, central entropy,
 and central temperature at bounce as well as the peak $\nu_e$-luminosity, the peak average
 $\nu_e$ energy, and the average $\nu_e$ energy prior to neutrino trapping, owing to
 variations of the progenitor model and electron-capture rates. Thirty two progenitors 
 were utilized from the WH07 model set of \citet{WOOSLEY2007} for producing the progenitor 
 bars (red) in the figure. Each bar of the electron-capture rate variations derives from 
 simulations where the rates have been systematically scaled by factors of 10, 4, 2, 0.5, 0.25,
 and 0.1. The horizontal tick represents the value of the reference simulation for the 
 tested Progenitor + EOS combination. The window ranges are chosen so that the progenitor 
 sensitivity bars are of equal size across each of the plotted parameters. 
 }
 \label{fig:sysranges}
\end{figure}

\subsubsection{Progenitor model sensitivity}
\label{progenitors}
In order to evaluate the significance of the electron-capture systematic sensitivity studies 
described so far, we have chosen to test against a study of the progenitor dependence 
of the core-collapse phase. Drawing from the larger set of progenitors from which 
the reference progenitors of the electron-capture study belong, the 2007
non-rotating solar-metallicity single-star model set from the stellar evolution code 
KEPLER~\citep{WOOSLEY2007} was utilized. This model set contains the presupernova configuration of 
32 stars ranging in zero-age-main-sequence (ZAMS) mass from 12 $M_\odot$ to 120 
$M_\odot$-- s12WH07 and s120WH07 respectively. 
Simulations of these progenitors exhibit a $\sim$3.5\% range of trapped lepton-fraction (0.288 - 0.298), 
a $\sim$4\% range of inner-core mass at bounce (0.473\,--\,0.491$M_\odot$), and a 
$\sim$9\% range of electron-neutrino peak luminosity ($5.19-5.65\cdot 10^{53}$ erg s$^{-1}$) 
during the neutrino flash occurring just after core bounce. 

Figure~\ref{fig:sysranges} compares the progenitor model and electron-capture rate 
dependence of several structural and neutrino quantities during collapse. 
The range of inner-core mass and peak $\nu_e$ luminosity seen from 
employing the WH07 progenitor model set are each approximately a 
factor of 5 smaller than the ranges seen from varying the 
electron-capture rates across all progenitor+EOS references. 
On the other hand, the range of central 
entropies and temperatures at bounce are comparable between the two 
sensitivity studies. The $\nu_e$ average energies 
just prior to neutrino trapping and during the deleptonization burst 
are also compared in Fig.~\ref{fig:sysranges}. The neutrinos emitted 
during the luminous burst just following core bounce are of  
higher energy than those emitted earlier because they arise primarily 
from electron capture on free protons. They also decouple from the core 
at a much hotter and denser neutrinosphere than prior to bounce, yielding 
higher energy neutrinos. In both of the sensitivity studies, variations of 
the electron-capture rates and of the initial stellar models, we find that the 
range of average neutrino energy during peak emission is comparable 
($\approx \pm$0.5 MeV). 

While captures on free protons contribute only marginally to deleptonization in the central zone, 
further out in the iron core, where the densities are lower (and $Y_e$'s are 
higher), electron captures on protons contribute to the deleptonization, especially in 
cases where we suppress electron captures on nuclei. The capture of 
electrons on these free protons produces neutrinos of a higher average energy, commensurate with the large 
spread seen in the bottom panel of Fig.~\ref{fig:sysranges} (which is taken when the central density is 
$3\times10^{11}\,$g\,cm$^{-3}$, but present from the onset of collapse). Another 
contribution to the energy spread is the systematic shift of electron captures to more neutron 
rich nuclei as the electron-capture rates are increased and the matter becomes more neutron rich. 
These neutron-rich nuclei have more negative Q-values, yielding lower energy neutrino emission.
Both of these effects result in a dispersion of average neutrino energies early on 
that is several factors larger than what is seen in the progenitor simulations.

Finally, we note that while the peak luminosity is only weakly dependent on the 
progenitor model, the post-bounce pre-explosion luminosity of all six neutrino species 
have strong progenitor dependences~\citep{OConnor2013}, and we find these pre-explosion luminosities 
are much less sensitive to the nuclear electron-capture rates comparatively--see panel (b) of 
Figure~\ref{fig:nulum}. The diverging of the luminosities seen at $t-t_b$ = 120 ms 
is due only to the difference in collapse times between the simulations 
which carries over to the evolution of the mass accretion rate after bounce.

\subsection{Monte-Carlo variations}
In addition to the possibility of systematic errors in the electron-capture rates, 
we also explore the effect of statistically distributed variations. Such an investigation 
is of great importance if the effect of an approximation such as Eq.~\ref{eq:ec} 
is to be understood. 
The main flaw in a continuous function for rate estimation across 
many nuclear species is the loss of structure, which would otherwise serve to 
statistically distribute the rates on a reaction by reaction basis (see Fig~\ref{fig:approx}). 
To study this effect we performed a Monte-Carlo (MC) variation of the electron-capture rates. 
Using an analytic description of the electron-capture rate distribtutions, 
such as a Gaussian or Poisson distribution, is likely to be inaccurate. 
Instead, we MC adjust the approximate rate 
for each species by adding to its log$_{10}(\lambda_{\mathrm{EC}})$ 
a value randomly chosen from a distribution created from the residuals of the tabulated 
rates and the approximate rates, i.e. 
\begin{align}
\textrm{log}_{10}(\lambda_{\mathrm{EC}}^{i,\textrm{table}}) - 
\textrm{log}_{10}(\lambda_{\mathrm{EC}}^{i,\textrm{Eq.1}})
\end{align}
where $i$ is an index running over all the tabulated reactions. 

In constructing this distribution, it is important also to preserve the Q-value dependence 
of the residuals that can be seen in Figure~\ref{fig:approx}b. We do this by seperating the 
residual distribution into subsets so that the reaction-rate residuals in each subset have 
similar Q-values. To do so we have chosen a Q-value binning of 2.5\,MeV, 
but have also tested binnings of 
5.0\,MeV and 10.0\,MeV which resolve the Q-dependence less, but have more counts 
per bin from which to sample. With this method we 
MC generate pseudo electron-capture rates that retain the Q-value dependence of Eq.~\ref{eq:ec}, 
but statistically distribute the approximate rate according to the variance of the 
rates calculated in the shell-model. Seven simulations for 
each binning were performed.

\begin{figure}
 \centering
 \includegraphics[width= 1. \columnwidth]{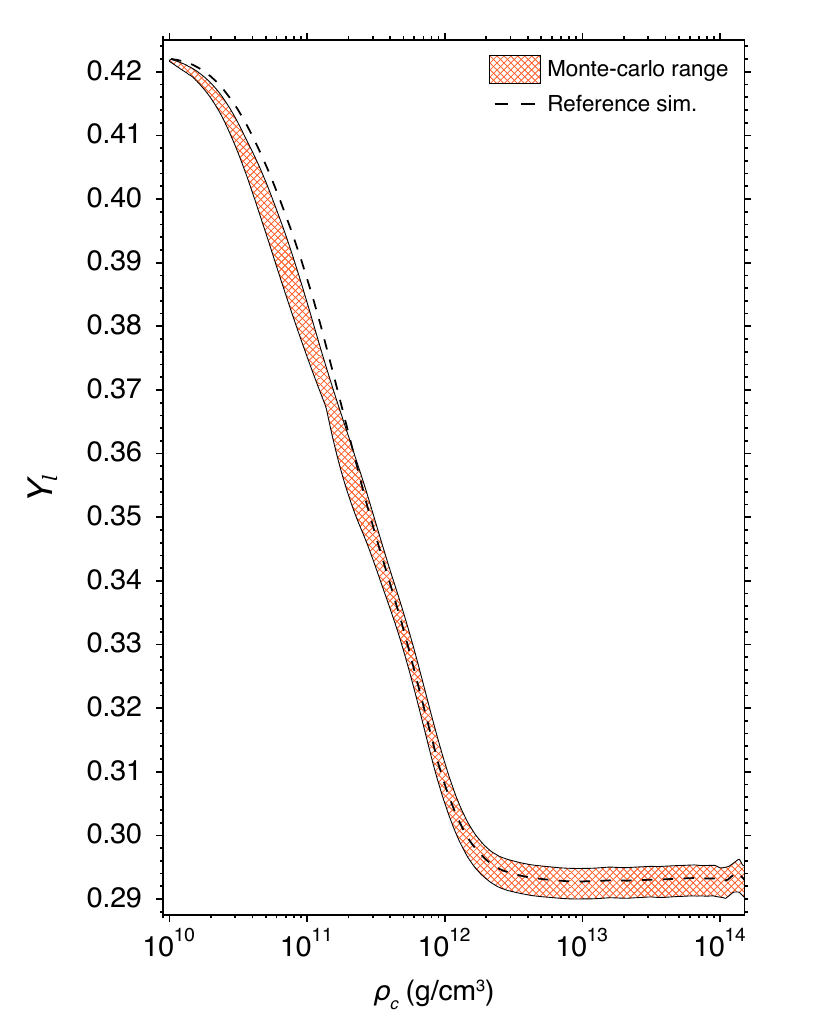}
 \caption{ Lepton fraction band as a function
 of central density for the Monte Carlo (MC) described in the text. The band represents
 data points from seven MC simulations based on the s15WW95+SFHo reference,
 where the band width represents the min to max range of $Y_l$.
 The rates are varied by drawing from the Q-value dependent residual function 
 shown in Figure~\ref{fig:approx} after binning it into 2.5 MeV 
 energy bins. This binning was chosen as it best tracks the Q-value dependence of 
 the residual distribution. 
 }
 \label{fig:ylep_stat}
\end{figure}

As mentioned before, at low densities the electron-capture rate depends strongly 
on the energy levels of the initial and final nuclides because the electron chemical 
potential is comparable to the excitation energies of the allowed Gamow-Teller transitions.
As the electron chemical potential increases, it encompasses a larger range 
of excitation energies which results in the electron-capture rate becoming 
sensitive primarily to the total strength. 
In the low-density case of Figure~\ref{fig:approx}a the approximation of Eq.~\ref{eq:ec} while appearing to 
decently reproduce the mean of the shell-model rates, actually has a mean approximately a factor 
of two lower than the tabulated rates. As the density increases, this difference 
between the mean electron-capture rate estimated by the approximation and the shell-model 
rates decrease. Thus, the approximation better reproduces the mean rate in the high density case 
of Figure~\ref{fig:approx}a. Because our MC simulations 
are based on residual distributions of the tabulated rates and the approximation, the 
average rate produced in each MC trial also has this bias. 

In Figure~\ref{fig:ylep_stat} we plot the min-to-max band representing the 
range of lepton fraction observed from all of the MC simulations. The band drawn corresponds to a 
2.5 MeV binning of the residual distributions from which the MC sampling was performed. 
For the reasons just described, the band has lower electron and lepton fractions than 
the reference at low densities, but becomes more statistically distributed around the reference at 
higher densities, near 5$\cdot$10$^{11}$g cm$^{-3}$.  
The lepton fraction band width varies from about a half percent initially, 
to its largest value of $\sim$2.5\% just before neutrino trapping, and then decreases 
back to $\sim$1.5\% before bounce. Altogether, we observe no significant impact on the core 
dynamics or the neutrino transport and therefore conclude that any statistically distributed 
scatter in the estimations of the electron-capture rates, such as those seen in Figure~\ref{fig:approx}, 
will likely not impact the models.

\section{Conclusion}
\label{conclusion}
Nuclear electron capture has long been understood to play an important role in 
the dynamics of core-collapse supernovae and large efforts have been undertaken to produce 
reliable estimates of electron-capture rates for astrophysical contexts. 
Although significant progress has been made in benchmarking theoretical electron-capture 
rates by comparison with charge-exchange experiments (especially using shell-model calculations) 
\citep{Cole2012}, large uncertainties remain for neutron-rich nuclei and nuclei beyond A=65.
Furthermore, sophisticated shell-model estimates for electron-capture rates exist 
only for a small subset of the large number of nuclei that contribute strongly. In this work 
we explore the implications of uncertainties in the electron-capture rate estimates for 
the core-collapse and early post-bounce phases of fully self consistent, general relativistic, 
core-collapse supernova simulations with comprehensive neutrino transport. 

\subsection{Most important nuclei}
For the reference 
simulation we calculate the contribution of each nucleus to core deleptonization and also perform a 
statistical resampling study, both of which identify those species whose rate should be 
known most precisely due to their significance in the simulations. With the given set of electron 
capture rates---from shell-model estimates to the approximate estimates of 
Eqs.~\ref{eq:ec} and~\ref{eq:nu}---we find that the simulations are most sensitive 
to neutron rich nuclei in the upper $pf$ and $pfg$/$sdg$-shells. 

Specifically, in these simulations nuclei near the A$\sim$80, N$\sim$50 closed neutron shell 
contribute the bulk of core deleptonization, and when removed from the simulations result in 
noticeable changes to the protoneutron star formation, with a significantly larger impact than 
when any other group of nuclei are removed. However, because sophisticated estimates from 
nuclear theory are not available for individual nuclei in this region, the electron-capture rates for 
these species have been accounted for in the past via simple averaging techniques and in 
this work via an approximation that has been fit to the LMP rate set. While this approximation 
reasonably reproduces the average electron capture rate for $sd$ and $pf$ shell nuclei near 
stability, rates for heavier neutron-rich nuclides will likely diverge from what is 
predicted by this parameterization. 

\subsection{Impact of uncertainties}
We evaluate the impact such uncertainties may have by varying 
the electron-capture rates for more than 6000 nuclei statistically, about the approximate 
prediction, and also systematically. On one hand, 
we find that statistical variations of electron-capture rates effect the overall dynamics and 
neutrino emission only weakly, producing marginal changes to the simulations. These findings 
indicate that the lack of structural variation that distributes the rate estimates from 
one species to the next is not crucial to the simulations. 

On the other hand, the average 
electron capture rate across a region of nuclei strongly determines the overall impact of 
those constituent nuclei. By systematically varying the electron-capture rates by factors 
between 10 and 0.1 we observe dramatic variations in the inner-core mass (+16/-4\,\%) and the 
electron-neutrino luminosity ($\pm$ 20\%) at and near bounce, respectively. 
Comparing with 32 simulations utilizing different 
progenitor models, we find that this range of inner-core mass and peak neutrino-luminosity 
is 5 times as large as that seen when varying the progenitor models.

We also find that the nuclear electron-capture rates 
are already large enough in the reference simulations that increasing them beyond their base 
values has a considerably smaller effect than decreasing them. This has compelling 
implications. Rates for A$\sim$80 nuclei near the N=50 shell gap, which have been shown in this work to 
be the primary contributors to the overall impact of electron captures during core collapse, 
may be overestimated by Eq.~\ref{eq:ec} due to Fermi-blocking at the closed neutron shell. 
Combined with a greater overall sensitivity to the systematic decrease in electron-capture rates, 
changes to the collapse and early post-bounce phases of the simulations may be as significant 
as those seen in this study if the current rates of these nuclei are found to be overestimated. 

\subsection{Goals for future studies}
For these reasons, it is important that experimental and theoretical efforts be aimed at nuclei 
which span the region on the chart of isotopes between stability and the neutron drip-line in 
both the $pf$ and $pfg/sdg$ model spaces, and further expand on the work that has been carried out for 
(near-)stable nuclei in the $pf$-shell. Since data from (n,p)-type charge-exchange experiments 
for nuclei in the $pfg/sdg$-shell and for neutron-rich nuclei in the $pf$ and $pfg/sdg$-shell are scarce, 
new experiments are required to obtain a sufficient set of data to benchmark current and future 
theoretical estimates. To this end, presently feasible  experiments on neutron-rich nuclei at and 
near stability with 60\textless A\textless 120  should add to 
the few cases that have been measured in this region.
With the higher beam intensities that will be available at next generation rare isotope 
facilities, future experimental programs should focus on the neutron-rich component of the 
primary electron-capture channel shown in Figure~\ref{fig:yedot}b. In particular, investigation 
of nuclei in the A$\sim$80 and N$\sim$50 region should take precedence, as changes to their electron-capture 
rates will significantly constrain the core-collapse dependence on nuclear electron-capture.

\section{Acknowledgements}
The authors would like to thank Gabriel Mart\'{i}nez-Pinedo and Raph Hix for valuable 
discussions, as well as the underlying rate tables from which much of this work 
was based. We would also like to thank Christian Ott for his support. CS would like to 
thank CITA for their hospitality and support while carrying out a portion of this research.
This material is based upon work supported by the National Science Foundation under Grants No. 
PHY-1430152 (JINA Center for the Evolution of the Elements) and No. PHY-1102511, by the 
Department of Energy (DOE) National Nuclear Security Administration under award number DE- NA0000979, 
and by NASA through Hubble Fellowship grant 51344.001-A awarded by the Space Telescope 
Science Institute, which is operated by the Association of Universities 
for Research in Astronomy, Inc., for NASA, under contract NAS5-26555,.
Some computations were performed on the Zwicky cluster at Caltech, which is supported by the 
Sherman Fairchild Foundation and by NSF award PHY-0960291.

\enlargethispage{3mm}
\pdfbookmark[1]{Bibliography}{biblio}
\bibliographystyle{apj}           
\bibliography{biblio}

\end{document}